\pdfoutput=1

\documentclass[12pt,a4paper]{article}

\usepackage{ifthen} 
\newboolean{pdflatex}
\setboolean{pdflatex}{true} 

\newboolean{articletitles}
\setboolean{articletitles}{true} 

\newboolean{uprightparticles}
\setboolean{uprightparticles}{false} 

\newboolean{inbibliography}
\setboolean{inbibliography}{false} 

\usepackage{microtype}
\usepackage{lineno}  
\usepackage{xspace} 

\usepackage{graphicx}  
\usepackage{color}
\usepackage{colortbl}
\graphicspath{{./figs/}} 

\usepackage{amsmath} 
\usepackage{amssymb}
\usepackage{amsfonts}
\usepackage{upgreek} 

\usepackage{multirow}

\newcommand*\patchAmsMathEnvironmentForLineno[1]{%
\expandafter\let\csname old#1\expandafter\endcsname\csname #1\endcsname
\expandafter\let\csname oldend#1\expandafter\endcsname\csname
end#1\endcsname
 \renewenvironment{#1}%
   {\linenomath\csname old#1\endcsname}%
   {\csname oldend#1\endcsname\endlinenomath}%
}
\newcommand*\patchBothAmsMathEnvironmentsForLineno[1]{%
  \patchAmsMathEnvironmentForLineno{#1}%
  \patchAmsMathEnvironmentForLineno{#1*}%
}
\AtBeginDocument{%
\patchBothAmsMathEnvironmentsForLineno{equation}%
\patchBothAmsMathEnvironmentsForLineno{align}%
\patchBothAmsMathEnvironmentsForLineno{flalign}%
\patchBothAmsMathEnvironmentsForLineno{alignat}%
\patchBothAmsMathEnvironmentsForLineno{gather}%
\patchBothAmsMathEnvironmentsForLineno{multline}%
}

\usepackage{hyperref}    
\usepackage[all]{hypcap} 

\def\lhcb {\mbox{LHCb}\xspace}

\def\babar  {\mbox{BaBar}\xspace}

\def\velo   {VELO\xspace}

\ifthenelse{\boolean{uprightparticles}}%
{
 \def\Pbeta       {\ensuremath{\upbeta}\xspace}
 \def\Pgamma      {\ensuremath{\upgamma}\xspace}

 \def\Peta        {\ensuremath{\upeta}\xspace}

 \def\Pmu         {\ensuremath{\upmu}\xspace}

 \def\Ppi         {\ensuremath{\uppi}\xspace}                 
                  
 \def\Prho        {\ensuremath{\uprho}\xspace}

 \def\Pchi        {\ensuremath{\upchi}\xspace}                 
 \def\Ppsi        {\ensuremath{\uppsi}\xspace}

 \def\PDelta      {\ensuremath{\Delta}\xspace}                 
 \def\PXi      {\ensuremath{\Xi}\xspace}                 
 \def\PLambda      {\ensuremath{\Lambda}\xspace}                 
 \def\PSigma      {\ensuremath{\Sigma}\xspace}                 
 \def\POmega      {\ensuremath{\Omega}\xspace}                 
 \def\PUpsilon      {\ensuremath{\Upsilon}\xspace}                 
 
 \def\PB      {\ensuremath{\mathrm{B}}\xspace}                 
                  
 \def\PD      {\ensuremath{\mathrm{D}}\xspace}

 \def\PJ      {\ensuremath{\mathrm{J}}\xspace}                 
 \def\PK      {\ensuremath{\mathrm{K}}\xspace}

 \def\Pb      {\ensuremath{\mathrm{b}}\xspace}                 
 \def\Pc      {\ensuremath{\mathrm{c}}\xspace}                 
 \def\Pd      {\ensuremath{\mathrm{d}}\xspace}

 \def\Ph      {\ensuremath{\mathrm{h}}\xspace}                 
 \def\Pi      {\ensuremath{\mathrm{i}}\xspace}

 \def\Pp      {\ensuremath{\mathrm{p}}\xspace}                 
 \def\Pq      {\ensuremath{\mathrm{q}}\xspace}                 
                  
 \def\Ps      {\ensuremath{\mathrm{s}}\xspace}

}
{
 \def\Pbeta       {\ensuremath{\beta}\xspace}
 \def\Pgamma      {\ensuremath{\gamma}\xspace}

 \def\Peta        {\ensuremath{\eta}\xspace}

 \def\Pmu         {\ensuremath{\mu}\xspace}

 \def\Ppi         {\ensuremath{\pi}\xspace}                 
                  
 \def\Prho        {\ensuremath{\rho}\xspace}

 \def\Pchi        {\ensuremath{\chi}\xspace}                 
 \def\Ppsi        {\ensuremath{\psi}\xspace}                 
                  
 \mathchardef\PDelta="7101
 \mathchardef\PXi="7104
 \mathchardef\PLambda="7103
 \mathchardef\PSigma="7106
 \mathchardef\POmega="710A
 \mathchardef\PUpsilon="7107
                  
 \def\PB      {\ensuremath{B}\xspace}                 
                  
 \def\PD      {\ensuremath{D}\xspace}

 \def\PJ      {\ensuremath{J}\xspace}                 
 \def\PK      {\ensuremath{K}\xspace}

 \def\Pb      {\ensuremath{b}\xspace}                 
 \def\Pc      {\ensuremath{c}\xspace}                 
 \def\Pd      {\ensuremath{d}\xspace}

 \def\Ph      {\ensuremath{h}\xspace}                 
 \def\Pi      {\ensuremath{i}\xspace}

 \def\Pp      {\ensuremath{p}\xspace}                 
 \def\Pq      {\ensuremath{q}\xspace}                 
                  
 \def\Ps      {\ensuremath{s}\xspace}

}

\def\mup        {\ensuremath{\Pmu^+}\xspace}
\def\mun        {\ensuremath{\Pmu^-}\xspace} 
\def\mumu       {\ensuremath{\Pmu^+\Pmu^-}\xspace}

\def\quark     {\ensuremath{\Pq}\xspace}
\def\quarkbar  {\ensuremath{\overline \quark}\xspace}

\def\dquark    {\ensuremath{\Pd}\xspace}

\def\squark    {\ensuremath{\Ps}\xspace}

\def\cquark    {\ensuremath{\Pc}\xspace}
\def\cquarkbar {\ensuremath{\overline \cquark}\xspace}

\def\bquark    {\ensuremath{\Pb}\xspace}

\def\pion  {\ensuremath{\Ppi}\xspace}
\def\piz   {\ensuremath{\pion^0}\xspace}

\def\pip   {\ensuremath{\pion^+}\xspace}
\def\pim   {\ensuremath{\pion^-}\xspace}
\def\pipm  {\ensuremath{\pion^\pm}\xspace}
\def\pimp  {\ensuremath{\pion^\mp}\xspace}

\def\kaon  {\ensuremath{\PK}\xspace}
  \def\Kbar  {\kern 0.2em\overline{\kern -0.2em \PK}{}\xspace}

\def\Kz    {\ensuremath{\kaon^0}\xspace}
\def\Kzb   {\ensuremath{\Kbar^0}\xspace}
\def\Kp    {\ensuremath{\kaon^+}\xspace}
\def\Km    {\ensuremath{\kaon^-}\xspace}
\def\Kpm   {\ensuremath{\kaon^\pm}\xspace}

\def\KS    {\ensuremath{\kaon^0_{\rm\scriptscriptstyle S}}\xspace} 
 
\def\Kstarz  {\ensuremath{\kaon^{*0}}\xspace}
\def\Kstarzb {\ensuremath{\Kbar^{*0}}\xspace}

\def\Kstarm  {\ensuremath{\kaon^{*-}}\xspace}

\newcommand{\etapr}{\ensuremath{\Peta^{\prime}}\xspace}

  \def\Dbar    {\kern 0.2em\overline{\kern -0.2em \PD}{}\xspace}
\def\D       {\ensuremath{\PD}\xspace}

\def\Dz      {\ensuremath{\D^0}\xspace}

\def\Dp      {\ensuremath{\D^+}\xspace}

\def\Dstarp  {\ensuremath{\D^{*+}}\xspace}

\def\Dsp     {\ensuremath{\D^+_\squark}\xspace}

\def\B       {\ensuremath{\PB}\xspace}
\def\Bbar    {\ensuremath{\kern 0.18em\overline{\kern -0.18em \PB}{}}\xspace}

\def\Bz      {\ensuremath{\B^0}\xspace}

\def\Bu      {\ensuremath{\B^+}\xspace}
\def\Bub     {\ensuremath{\B^-}\xspace}

\def\Bm      {\ensuremath{\Bub}\xspace}

\def\Bd      {\ensuremath{\B^0}\xspace}
\def\Bs      {\ensuremath{\B^0_\squark}\xspace}

\def\jpsi     {\ensuremath{{\PJ\mskip -3mu/\mskip -2mu\Ppsi\mskip 2mu}}\xspace}

\def\chiczero {\ensuremath{\Pchi_{\cquark 0}}\xspace}

  \def\Y#1S{\ensuremath{\PUpsilon{(#1S)}}\xspace}

\def\proton      {\ensuremath{\Pp}\xspace}

\def\L {\ensuremath{\PLambda}\xspace}
\def\Lbar {\ensuremath{\kern 0.1em\overline{\kern -0.1em\PLambda}}\xspace}

\def\Lb      {\ensuremath{\L^0_\bquark}\xspace}

\def\Lc      {\ensuremath{\L^+_\cquark}\xspace}

\def\BF         {{\ensuremath{\cal B}\xspace}}

\newcommand{\decay}[2]{\ensuremath{#1\!\to #2}\xspace}         

\def\to                 {\ensuremath{\rightarrow}\xspace}

\def\eps   {\ensuremath{\varepsilon}\xspace}

\def\CP                {\ensuremath{C\!P}\xspace}

\newcommand{\betas}{\ensuremath{\beta_{\squark}}\xspace}

\def\AT#1     {\ensuremath{A_{\mathrm{T}}^{#1}}\xspace}           

\def\C#1      {\ensuremath{\mathcal{C}_{#1}}\xspace}                       
\def\Cp#1     {\ensuremath{\mathcal{C}_{#1}^{'}}\xspace}                    
\def\Ceff#1   {\ensuremath{\mathcal{C}_{#1}^{\mathrm{(eff)}}}\xspace}        
\def\Cpeff#1  {\ensuremath{\mathcal{C}_{#1}^{'\mathrm{(eff)}}}\xspace}       
\def\Ope#1    {\ensuremath{\mathcal{O}_{#1}}\xspace}                       
\def\Opep#1   {\ensuremath{\mathcal{O}_{#1}^{'}}\xspace}                    

\newcommand{\tev}{\ifthenelse{\boolean{inbibliography}}{\ensuremath{~T\kern -0.05em eV}\xspace}{\ensuremath{\mathrm{\,Te\kern -0.1em V}}\xspace}}
\newcommand{\gev}{\ensuremath{\mathrm{\,Ge\kern -0.1em V}}\xspace}
\newcommand{\mev}{\ensuremath{\mathrm{\,Me\kern -0.1em V}}\xspace}
\newcommand{\kev}{\ensuremath{\mathrm{\,ke\kern -0.1em V}}\xspace}
\newcommand{\ev}{\ensuremath{\mathrm{\,e\kern -0.1em V}}\xspace}
\newcommand{\gevc}{\ensuremath{{\mathrm{\,Ge\kern -0.1em V\!/}c}}\xspace}
\newcommand{\mevc}{\ensuremath{{\mathrm{\,Me\kern -0.1em V\!/}c}}\xspace}
\newcommand{\gevcc}{\ensuremath{{\mathrm{\,Ge\kern -0.1em V\!/}c^2}}\xspace}
\newcommand{\gevgevcccc}{\ensuremath{{\mathrm{\,Ge\kern -0.1em V^2\!/}c^4}}\xspace}
\newcommand{\mevcc}{\ensuremath{{\mathrm{\,Me\kern -0.1em V\!/}c^2}}\xspace}

\def\mm   {\ensuremath{\rm \,mm}\xspace}

\def\mum  {\ensuremath{\,\upmu\rm m}\xspace}

\def\invfb   {\ensuremath{\mbox{\,fb}^{-1}}\xspace}

\newcommand{\chisq}{\ensuremath{\chi^2}\xspace}

\newcommand{\chisqip}{\ensuremath{\chi^2_{\rm IP}}\xspace}
\newcommand{\chisqvs}{\ensuremath{\chi^2_{\rm VS}}\xspace}
\newcommand{\chisqvtx}{\ensuremath{\chi^2_{\rm vtx}}\xspace}

\def\gsim{{~\raise.15em\hbox{$>$}\kern-.85em
          \lower.35em\hbox{$\sim$}~}\xspace}
\def\lsim{{~\raise.15em\hbox{$<$}\kern-.85em
          \lower.35em\hbox{$\sim$}~}\xspace}

\def\sPlot{\mbox{\em sPlot}}

\def\ptot       {\mbox{$p$}\xspace}
\def\pt         {\mbox{$p_{\rm T}$}\xspace}
\def\et         {\mbox{$E_{\rm T}$}\xspace}

\def\dllkpi     {\ensuremath{\mathrm{DLL}_{\kaon\pion}}\xspace}
\def\dllppi     {\ensuremath{\mathrm{DLL}_{\proton\pion}}\xspace}

\def\evtgen     {\mbox{\textsc{EvtGen}}\xspace}

\def\geant      {\mbox{\textsc{Geant4}}\xspace}

\def\photos     {\mbox{\textsc{Photos}}\xspace}

\def\pythia     {\mbox{\textsc{Pythia}}\xspace}

\def\tell1  {TELL1\xspace}
\def\ukl1   {UKL1\xspace}

\newcommand{\eg}{\mbox{\itshape e.g.}\xspace}

\usepackage{cite} 
\usepackage{mciteplus}

\def\dllpk     {\ensuremath{\mathrm{DLL}_{\proton\kaon}}\xspace}

\ifthenelse{\boolean{uprightparticles}}%
{\def\Prho      {\ensuremath{\uprho}\xspace}
 
}
{\def\Prho      {\ensuremath{\rho}\xspace}
 
}
\def\rhoz   {\ensuremath{\Prho^0}\xspace}

\def\had  {\ensuremath{\Ph}\xspace}
\def\hadp  {\ensuremath{\Ph^+}\xspace}
\def\hadm  {\ensuremath{\Ph^-}\xspace}

\def\hadprimm {\ensuremath{\had^{\prime-}}\xspace}

\def\pipi  {\ensuremath{\pip\pim}\xspace}
\def\KpKm  {\ensuremath{\Kp \kern -0.16em \Km}\xspace}
\def\KzKzb {\ensuremath{\Kz \kern -0.16em \Kzb}\xspace}

\def\Bdz      {\ensuremath{\Bd}\xspace}
\def\Bsz      {\ensuremath{\Bs}\xspace}
\def\Bdsz      {\ensuremath{\Bz_{(\squark)}}\xspace}

\def\mumu      {\ensuremath{\mup\mun}\xspace}

\def\pp    {\ensuremath{\proton\proton}\xspace}

\def\BdtoKzKK   {\decay{\Bd}{\Kz \Kp \Km}}
\def\BdtoKzPiPi   {\decay{\Bd}{\Kz \pip \pim}}
\def\BdtoKzKPi   {\decay{\Bd}{\Kz \Kpm \pimp}}

\def\BstoKzKK   {\decay{\Bs}{\Kz \Kp \Km}}
\def\BstoKzPiPi   {\decay{\Bs}{\Kz \pip \pim}}
\def\BstoKzKPi   {\decay{\Bs}{\Kz \Kpm \pimp}}

\def\BdtoKsKK   {\decay{\Bd}{\KS \Kp \Km}}
\def\BdtoKsPiPi   {\decay{\Bd}{\KS \pip \pim}}
\def\BdtoKsKPi   {\decay{\Bd}{\KS \Kpm \pimp}}

\def\BstoKsKK   {\decay{\Bs}{\KS \Kp \Km}}
\def\BstoKsPiPi   {\decay{\Bs}{\KS \pip \pim}}
\def\BstoKsKPi   {\decay{\Bs}{\KS \Kpm \pimp}}

\def\BdstoKsKK   {\decay{\Bdsz}{\KS \Kp \Km}}
\def\BdstoKsPiPi   {\decay{\Bdsz}{\KS \pip \pim}}
\def\BdstoKsKPi   {\decay{\Bdsz}{\KS \Kpm \pimp}}

\def\Kshhp{\ensuremath{\KS \hadp \hadprimm}\xspace}

\def\KsPiPi{\ensuremath{\KS \pip \pim}\xspace}
\def\KsKPi{\ensuremath{\KS \Kpm \pimp}\xspace}
\def\KsKK{\ensuremath{\KS \Kp \Km}\xspace}

\def\BstoKshhp   {\decay{\Bs}{\KS \hadp \hadprimm}}

\def\BdstoKshhp   {\decay{\Bdsz}{\KS \hadp \hadprimm}}

\def\BstoKPiPiz {\decay{\Bs}{\Km \pip \piz}}

\def\BdtoetapKs   {\decay{\Bdz}{\etapr (\to \rhoz \gamma) \KS}}

\def\ButoKsPiPiPi   {\decay{\Bu}{\KS \pip \pim \pip}}

\def\BstoKstKstbartoKsPizKPi   {\decay{\Bsz}{\Kstarz (\to \KS \piz) \Kstarzb (\to \Km \pip)}}

\def\btoqqbars {\decay{\bquark}{\quark\quarkbar\squark}}

\def\btoccbars {\decay{\bquark}{\cquark\cquarkbar\squark}}

\def\LL   {Long\xspace}
\def\DD   {Downstream\xspace}

\newcommand{\Br}[1]{\ensuremath{\BF\left(#1\right)}\xspace}

\def\fsfdinline {\ensuremath{f_{\squark}/f_{\dquark}}\xspace}

\newcommand{\tab}[1]{Table~\ref{tab : #1}}

\newcommand{\fig}[1]{Fig.~\ref{fig : #1}}

\newcommand{\figstwo}[2]{Figs.~\ref{fig : #1} and~\ref{fig : #2}}

\def\argus {ARGUS\xspace}

\begin{document}

\renewcommand{\thefootnote}{\fnsymbol{footnote}}
\setcounter{footnote}{1}


\begin{titlepage}
\pagenumbering{roman}

\vspace*{-1.5cm}
\centerline{\large EUROPEAN ORGANIZATION FOR NUCLEAR RESEARCH (CERN)}
\vspace*{1.5cm}
\hspace*{-0.5cm}
\begin{tabular*}{\linewidth}{lc@{\extracolsep{\fill}}r}
\ifthenelse{\boolean{pdflatex}}
{\vspace*{-2.7cm}\mbox{\!\!\!\includegraphics[width=.14\textwidth]{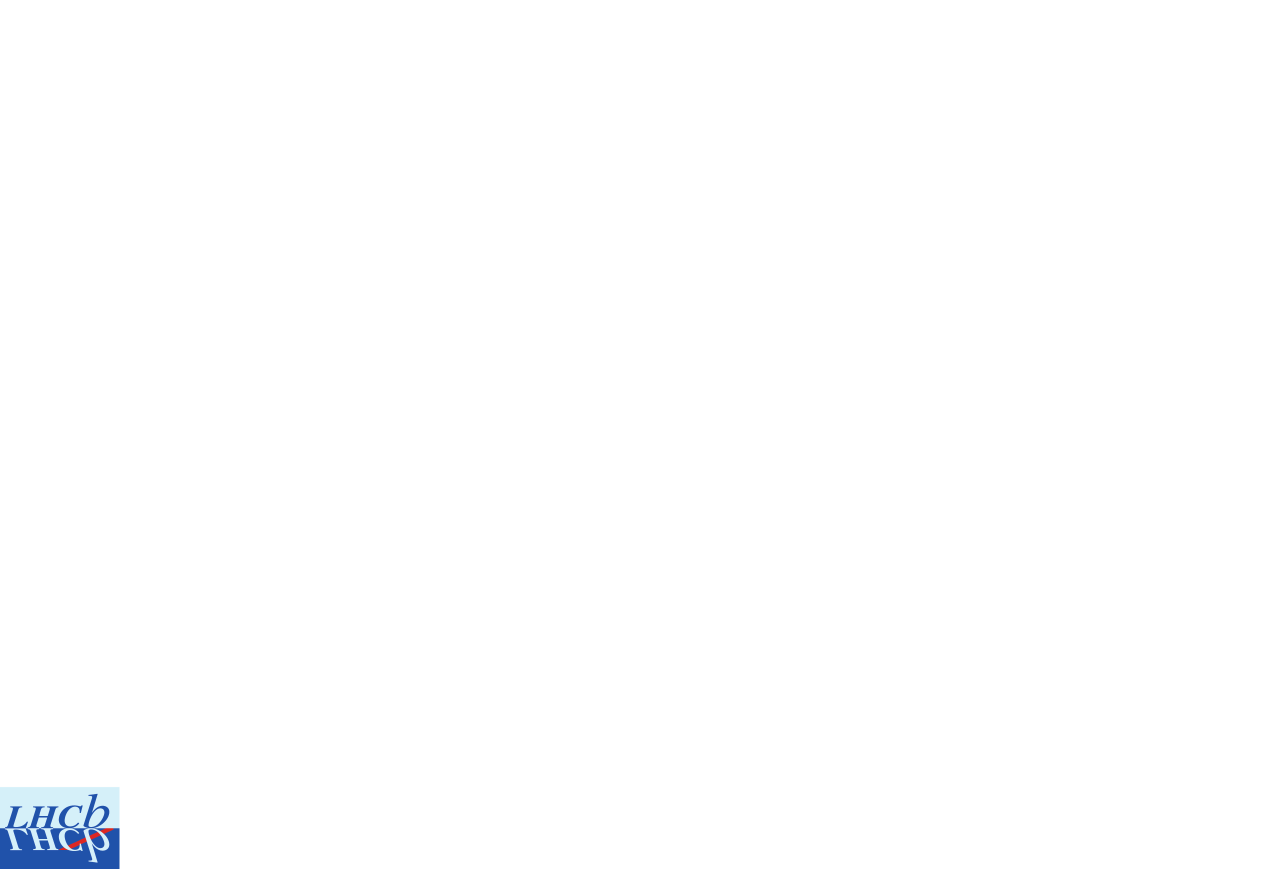}} & &}%
{\vspace*{-1.2cm}\mbox{\!\!\!\includegraphics[width=.12\textwidth]{figs/lhcb-logo.eps}} & &}%
\\
 & & CERN-PH-EP-2013-139 \\  
 & & LHCb-PAPER-2013-042 \\  
 & & July 29, 2013 \\ 
 & & \\
\end{tabular*}

\vspace*{0.70cm}

{\bf\boldmath\Large
\begin{center}
Study of 
$B_{\scriptscriptstyle (s)}^0 \to K_{\rm \scriptscriptstyle S}^0 h^{+} h^{\prime -}$
decays with
first observation of
$B_{\scriptscriptstyle s}^0 \to K_{\rm \scriptscriptstyle S}^0 K^{\pm} \pi^{\mp}$
and
$B_{\scriptscriptstyle s}^0 \to K_{\rm \scriptscriptstyle S}^0 \pi^{+} \pi^{-}$
\end{center}
}

\vspace*{0.40cm}

\begin{center}
The LHCb collaboration\footnote{Authors are listed on the following pages.}
\end{center}

\vspace{\fill}

\begin{abstract}
  \noindent
  A search for charmless three-body decays of $B^0$ and
  $B_{\scriptscriptstyle s}^0$ mesons with a
  $K_{\rm \scriptscriptstyle S}^0$ meson in the final state
  is performed using the $pp$ collision data,
  corresponding to an integrated luminosity of $1.0\mbox{\,fb}^{-1}$,
  collected at a centre-of-mass energy of $7\mathrm{\,Te\kern -0.1em V}$
  recorded by the LHCb experiment.
  Branching fractions of the
  $B_{\scriptscriptstyle (s)}^0 \to K_{\rm \scriptscriptstyle S}^0 h^{+} h^{\prime -}$
  decay modes ($h^{(\prime)} = \pi, K$), relative to the well measured
  $B^0 \to K_{\rm \scriptscriptstyle S}^0 \pi^{+} \pi^{-}$
  decay, are obtained.
  First observation of the decay modes
  $B_s^0 \to K_{\rm \scriptscriptstyle S}^0 K^{\pm} \pi^{\mp}$ 
  and
  $B_s^0 \to K_{\rm \scriptscriptstyle S}^0 \pi^{+} \pi^{-}$ 
  and confirmation of the decay
  $B^0 \to K_{\rm \scriptscriptstyle S}^0 K^{\pm} \pi^{\mp}$
  are reported.
  The following relative branching fraction measurements or limits are obtained
  \begin{eqnarray*}
  \nonumber
  \frac{{\cal B}(B^0   \to K_{\rm \scriptscriptstyle S}^0 K^{\pm}   \pi^{\mp})}{{\cal B}(B^0 \to K_{\rm \scriptscriptstyle S}^0 \pi^{+} \pi^{-})} &=& 0.128 \pm 0.017 \, ({\rm stat.}) \pm 0.009 \, ({\rm syst.}) \,, \\
  \nonumber
  \frac{{\cal B}(B^0   \to K_{\rm \scriptscriptstyle S}^0 K^{+}     K^{-}    )}{{\cal B}(B^0 \to K_{\rm \scriptscriptstyle S}^0 \pi^{+} \pi^{-})} &=& 0.385 \pm 0.031 \, ({\rm stat.}) \pm 0.023 \, ({\rm syst.}) \,, \\
  \nonumber
  \frac{{\cal B}(B_s^0 \to K_{\rm \scriptscriptstyle S}^0 \pi^{+}   \pi^{-}  )}{{\cal B}(B^0 \to K_{\rm \scriptscriptstyle S}^0 \pi^{+} \pi^{-})} &=& 0.29\phantom{0}  \pm 0.06\phantom{0}  \, ({\rm stat.}) \pm 0.03\phantom{0}  \, ({\rm syst.}) \pm 0.02 \, (f_s/f_d) \,, \\
  \nonumber
  \frac{{\cal B}(B_s^0 \to K_{\rm \scriptscriptstyle S}^0 K^{\pm}   \pi^{\mp})}{{\cal B}(B^0 \to K_{\rm \scriptscriptstyle S}^0 \pi^{+} \pi^{-})} &=& 1.48\phantom{0}  \pm 0.12\phantom{0}  \, ({\rm stat.}) \pm 0.08\phantom{0}  \, ({\rm syst.}) \pm 0.12 \, (f_s/f_d) \,, \\
  \nonumber    
  \frac{{\cal B}(B_s^0 \to K_{\rm \scriptscriptstyle S}^0 K^{+}     K^{-}    )}{{\cal B}(B^0 \to K_{\rm \scriptscriptstyle S}^0 \pi^{+} \pi^{-})} &\in& [0.004;0.068]  \; {\rm at \;\; 90\% \; CL}  \,.
  \end{eqnarray*}  
\end{abstract}

\vspace*{1.0cm}

\begin{center}
  Submitted to JHEP
\end{center}

\vspace{\fill}

{\footnotesize 
\centerline{\copyright~CERN on behalf of the \lhcb collaboration, license \href{http://creativecommons.org/licenses/by/3.0/}{CC-BY-3.0}.}}
\vspace*{2mm}

\end{titlepage}


\newpage
\setcounter{page}{2}
\mbox{~}
\newpage

\centerline{\large\bf LHCb collaboration}
\begin{flushleft}
\small
R.~Aaij$^{40}$, 
B.~Adeva$^{36}$, 
M.~Adinolfi$^{45}$, 
C.~Adrover$^{6}$, 
A.~Affolder$^{51}$, 
Z.~Ajaltouni$^{5}$, 
J.~Albrecht$^{9}$, 
F.~Alessio$^{37}$, 
M.~Alexander$^{50}$, 
S.~Ali$^{40}$
% ...

{\footnotesize \it
% ...
$ ^{5}$Clermont Universit\'{e}, Universit\'{e} Blaise Pascal, CNRS/IN2P3, LPC, Clermont-Ferrand, France\\
$ ^{6}$CPPM, Aix-Marseille Universit\'{e}, CNRS/IN2P3, Marseille, France\\
% ...
$ ^{9}$Fakult\"{a}t Physik, Technische Universit\"{a}t Dortmund, Dortmund, Germany\\
% ...
$ ^{36}$Universidad de Santiago de Compostela, Santiago de Compostela, Spain\\
$ ^{37}$European Organization for Nuclear Research (CERN), Geneva, Switzerland\\
% ...
$ ^{40}$Nikhef National Institute for Subatomic Physics, Amsterdam, The Netherlands\\
% ...
$ ^{45}$H.H. Wills Physics Laboratory, University of Bristol, Bristol, United Kingdom\\
% ...
$ ^{50}$School of Physics and Astronomy, University of Glasgow, Glasgow, United Kingdom\\
$ ^{51}$Oliver Lodge Laboratory, University of Liverpool, Liverpool, United Kingdom\\
% ...
}
\end{flushleft}

\cleardoublepage


\renewcommand{\thefootnote}{\arabic{footnote}}
\setcounter{footnote}{0}


\pagestyle{plain} 
\setcounter{page}{1}
\pagenumbering{arabic}

%

\section{Introduction}
\label{sec:intro}

The study of the charmless three-body decays of neutral \B mesons to final
states including a \KS meson, namely \BdstoKsPiPi, \BdstoKsKPi and \BdstoKsKK,
has a number of theoretical applications.\footnote{Unless stated otherwise, charge conjugated modes are implicitly included throughout the paper.}
The decays \BdtoKsPiPi and \BdtoKsKK are dominated by \btoqqbars ($q = u,d,s$)
loop transitions.
Mixing-induced \CP asymmetries in such decays are predicted to be approximately
equal to those in \btoccbars transitions, \eg $\Bd\to\jpsi\KS$, by the
Cabibbo-Kobayashi-Maskawa mechanism~\cite{Cabibbo:1963yz,Kobayashi:1973fv}.
However, the loop diagrams that dominate the charmless decays can have
contributions from new particles in several extensions of the Standard Model,
which could introduce additional weak phases~\cite{Buchalla:2005us,Grossman:1996ke,London:1997zk,Ciuchini:1997zp}.
A time-dependent analysis of the three-body Dalitz plot allows measurements of
the mixing-induced \CP-violating
phase~\cite{Dalseno:2008wwa,Aubert:2009me,Nakahama:2010nj,Lees:2012kxa}. 
 The current experimental measurements of \btoqqbars decays~\cite{HFAG} show
fair agreement with the results from \btoccbars decays (measuring the weak
phase \Pbeta) for each of the scrutinised \CP eigenstates.
There is, however, a global trend towards lower values than the weak phase
measured from \btoccbars decays.
The interpretation of this deviation is made complicated by QCD
corrections, which depend on the final state~\cite{Silvestrini:2007yf} and
are difficult to handle.
An analogous extraction of the mixing-induced \CP-violating phase in the
\Bs system will, with a sufficiently large dataset, also be possible with
the \BstoKsKPi decay, which can be compared with that from, \eg
$\Bs\to\jpsi\phi$.

Much recent theoretical and experimental activity has focused on the
determination of the CKM angle \Pgamma from $B \to K\pi\pi$ decays, using
and refining the methods proposed in Refs.~\cite{Ciuchini:2006kv,Gronau:2006qn}.
The recent experimental results from \babar~\cite{BABAR:2011ae} demonstrate the
feasibility of the method, albeit with large statistical uncertainties.
The decay \BstoKsPiPi is of particular interest for this effort.
Indeed, the ratio of the amplitudes of the isospin-related mode \BstoKPiPiz and
its charge conjugate exhibits a direct dependence on the mixing-induced
\CP-violating phase, which would be interpreted in the Standard Model as
$(\betas + \gamma)$.
Unlike the equivalent \Bd decays, the \Bs decays are dominated by tree
amplitudes and the contributions from electroweak penguin diagrams are expected
to be negligible, yielding a theoretically clean extraction of
\Pgamma~\cite{Ciuchini:2006st} provided that the strong phase
can be determined from other measurements.
The shared intermediate states between \BstoKPiPiz and \BstoKsPiPi (specifically
$\Kstarm\pip$) offer that possibility, requiring an analysis of the
\BstoKsPiPi Dalitz plot.

At LHCb, the first step towards this physics programme is to establish the
signals of all the decay modes.
In particular, the decay modes \BstoKshhp ($\had^{(\prime)} = \pi, K$) are all
unobserved and the observation of \BdtoKsKPi by
\babar~\cite{delAmoSanchez:2010ur} is so far unconfirmed.
In this paper the results of an analysis of all six \BdstoKshhp decay modes are presented.
The branching fractions of the decay modes relative to that of \BdtoKsPiPi are measured when 
the significance of the signals allow it, otherwise confidence intervals are quoted.
Time-integrated branching fractions are computed, implying a non-trivial
comparison of the \Bd and \Bs decays at amplitude level~\cite{DeBruyn:2012wj}.

\section{Detector and dataset}
\label{sec:Detector}

The measurements described in this paper are performed with data,
corresponding to an integrated luminosity of 1.0\invfb,
from $7\tev$ centre-of-mass \pp collisions,
collected with the \lhcb detector during 2011. 
Samples of simulated events are used to estimate the efficiency
of the selection requirements, to investigate possible sources of
background contributions, and to model the event distributions in the likelihood fit.
In the simulation, $pp$ collisions are generated using
\pythia~6.4~\cite{Sjostrand:2006za} with a specific \lhcb
configuration~\cite{LHCb-PROC-2010-056}.  Decays of hadronic particles
are described by \evtgen~\cite{Lange:2001uf}, in which final state
radiation is generated using \photos~\cite{Golonka:2005pn}. The
interaction of the generated particles with the detector and its
response are implemented using the \geant
toolkit~\cite{Allison:2006ve, *Agostinelli:2002hh} as described in
Ref.~\cite{LHCb-PROC-2011-006}.

The \lhcb detector~\cite{Alves:2008zz} is a single-arm forward
spectrometer covering the \mbox{pseudorapidity} range $2<\eta <5$,
designed for the study of particles containing \bquark or \cquark
quarks. The detector includes a high-precision tracking system
consisting of a silicon-strip vertex detector (VELO) surrounding the $pp$
interaction region, a large-area silicon-strip detector located
upstream of a dipole magnet with a bending power of about
$4{\rm\,Tm}$, and three stations of silicon-strip detectors and straw
drift tubes placed downstream.
The combined tracking system provides a momentum measurement with relative
uncertainty that varies from 0.4\% at 5\gevc to 0.6\% at 100\gevc, and
impact parameter resolution of 20\mum for tracks with high transverse
momentum.
Charged hadrons are identified using two ring-imaging Cherenkov (RICH)
detectors~\cite{LHCb-DP-2012-003}.
Photon, electron and hadron candidates are identified by a calorimeter
system consisting of scintillating-pad and preshower detectors, an
electromagnetic calorimeter and a hadronic calorimeter.
Muons are identified by a system composed of alternating layers of iron and
multiwire proportional chambers.

\section{Trigger and event selection}
\label{sec:EventSelection}

The trigger~\cite{LHCb-DP-2012-004} consists of a hardware stage, based on
information from the calorimeter and muon systems, followed by a software
stage, which applies a full event reconstruction.
To remove events with large occupancies, a requirement is made at the
hardware stage on the number of hits in the scintillating-pad detector.
The hadron trigger at the hardware stage also requires that there is at
least one candidate with transverse energy $\et > 3.5\gev$.
In the offline selection, candidates are separated into two categories
based on the hardware trigger decision.
The first category are triggered by particles from candidate signal decays
that have an associated cluster in the calorimeters above the threshold,
while the second category are triggered independently of the particles
associated with the signal decay.
Events that do not fall into either of these categories are not used in the
subsequent analysis. 

The software trigger requires a two-, three- or four-track secondary vertex
with a high sum of the transverse momentum, \pt, of the tracks and
significant displacement from the primary $pp$ interaction vertices~(PVs).
At least one track should have $\pt > 1.7\gevc$ and \chisqip with respect
to any primary interaction greater than 16, where \chisqip is defined as
the difference in \chisq of a given PV reconstructed with and without the
considered track. A multivariate algorithm~\cite{BBDT} is used for the
identification of secondary vertices consistent with the decay of a \bquark
hadron.

The events passing the trigger requirements are then filtered in two stages.
Initial requirements are applied to further reduce the size of the data
sample, before a multivariate selection is implemented.
In order to minimise the variation of the selection efficiency over the Dalitz
plot it is necessary to place only loose requirements on the momenta of the
daughter particles.
As a consequence, selection requirements on topological variables such as the
flight distance of the \B candidate or the direction of its momentum vector are
used as the main discriminants.

The \KS candidates are reconstructed in the \pipi final state.
Approximately two thirds of the reconstructed \KS mesons decay downstream of
the \velo.
Since those \KS candidates decaying within the \velo, and those that have
information only from the tracking stations, differ in their reconstruction
and selection, they are separated into two categories labelled ``\LL'' and
``\DD''\!, respectively.
The pions that form the \KS candidates are required to have momentum
$\ptot>2\gevc$ and \chisqip with respect to any PV greater than 9 (4) for
\LL (\DD) \KS candidates.
The \KS candidates are then required to form a vertex with $\chisqvtx<12$
and to have invariant mass within 20\mevcc (30\mevcc) of the nominal \KS
mass~\cite{PDG2012} for \LL (\DD) candidates.
The square of the separation of the \KS vertex from the PV divided by the
associated uncertainty (\chisqvs) must be greater than $80$ ($50$) for \LL
(\DD) candidates.
\DD \KS candidates are required, in addition, to have momentum $\ptot>6\gevc$.

The \B candidates are formed by combining the \KS candidates with two
oppositely charged tracks.
Selection requirements, common to both the \LL and \DD categories, are
based on the topology and kinematics of the \B candidate.
The charged \B-meson daughters are required to have $\ptot<100\gevc$, a
momentum beyond which there is little pion/kaon discrimination.
The scalar sum of the three daughters' transverse momenta must be greater
than 3\gevc, and at least two of the daughters must have $\pt>0.8\gevc$.
The impact parameter (IP) of the \B-meson daughter with the largest \pt is
required to be greater than 0.05\mm relative to the PV associated to
the \B candidate.
The \chisq of the distance of closest approach of any two daughters must be
less than 5.
The \B candidates are then required to form a vertex separated from any PV
by at least 1\mm and that has $\chisqvtx<12$ and $\chisqvs>50$.
The difference in \chisqvtx when adding any track must be greater than
4.
The candidates must have $\pt>1.5\gevc$ and invariant mass within
the range $4779 < m_{\Kshhp} < 5866 \mevcc$.
The cosine of the angle between the reconstructed momentum of the \B meson
and its direction of flight (pointing angle) is required to be greater than
0.9999.
The candidates are further required to have a minimum \chisqip with respect
to all PVs less than 4.
Finally, the separation of the \KS and \B vertices in the positive $z$
direction\footnote{The $z$ axis points along the beam line from the
interaction region through the LHCb detector.}
must be greater than 30\mm.

Multivariate discriminants based on a boosted decision tree~(BDT)~\cite{Breiman}
with the AdaBoost algorithm~\cite{AdaBoost} have been designed in order to
complete the selection of the signal events and to further reject
combinatorial backgrounds.
Simulated \BdstoKsPiPi events and upper mass sidebands,
$5420 < m_{\KsPiPi} < 5866 \mevcc$,
in the data are used as the signal and background training samples,
respectively.
The samples of events in each of the \LL and \DD \KS categories are further
subdivided into two equally-sized subsamples.  Each subsample is then used
to train an independent discriminant.
In the subsequent analysis the BDT trained on one subsample of a given \KS
category is used to select events from the other subsample, in order to
avoid bias.
The input variables for the BDTs are the \pt, $\eta$, \chisqip, \chisqvs,
pointing angle and \chisqvtx of the \B candidate; the sum \chisqip of the
\hadp and \hadm; the \chisqip, \chisqvs and \chisqvtx of the \KS candidate.

The selection requirement placed on the output of the BDTs is independently
optimised for events containing \KS candidates reconstructed in either \DD
or \LL categories.
Two different figures of merit are used to optimise the selection
requirements, depending on whether the decay mode in question is favoured
or suppressed.
If favoured, the following is used
\begin{equation}
{\cal Q}_1 = \frac{{\rm S}}{\sqrt{{\rm S}+{\rm B}}} \,,
\end{equation}
where $\rm S$ ($\rm B$) represents the number of expected signal
(combinatorial background) events for a given selection.
The value of $\rm S$ is estimated based on the known branching fractions
and efficiencies, while $\rm B$ is calculated by fitting the sideband above
the signal region and extrapolating into the signal region.
If the mode is suppressed, an alternative figure of
merit~\cite{Punzi:2003bu} is used
\begin{equation}
{\cal Q}_2 = \frac{\eps_{\rm sig}}{\frac{a}{2} + \sqrt{\rm B}} \,,
\end{equation}
where the signal efficiency ($\eps_{\rm sig}$) is estimated from
the signal simulation.
The value $a=5$ is used in this analysis, which corresponds to optimising
for $5\sigma$ significance to find the decay.
This second figure of merit results in a more stringent requirement than the
first.  Hence, the requirements optimised with each figure of merit will
from here on be referred to as the loose and tight BDT requirements,
respectively.

The fraction of selected events containing more than one candidate is at
the percent level.  The candidate to be retained in each event is chosen
arbitrarily.

A number of background contributions consisting of fully reconstructed \B meson decays into
two-body $\D\had$ or $\cquark\cquarkbar\KS$ combinations, result in a
\Kshhp final state and hence are, in terms of their \B candidate invariant mass
distribution, indistinguishable from signal candidates.
The decays of \Lb baryons to $\Lc\had$ with \decay{\Lc}{\proton\KS} also
peak under the signal when the proton is misidentified.
Therefore, the following \D, \Lc and charmonia decays are explicitly
reconstructed under the relevant particle hypotheses and vetoed in all the
spectra:
$\Dz \to \Km\pip$,
$\Dz \to \pip\pim$,
$\Dz \to \Kp\Km$,
$\Dp \to \KS\Kp$,
$\Dp \to \KS\pip$,
$\Dsp \to \KS\Kp$,
$\Dsp \to \KS\pip$,
and
$\Lc \to \proton\KS$.
Additional vetoes on charmonium resonances,
$\jpsi \to \pipi, \mumu, \KpKm$ and 
$\chiczero \to \pipi, \mumu, \KpKm$, are applied to remove the handful of fully reconstructed and well identified peaking \decay{\Bdsz}{\left(\jpsi,\chiczero\right)\KS} decays.
The veto for each reconstructed charm (charmonium) state $R$,
$\left|m-m_R\right| <  30 \; (48) \mevcc$,
is defined around the world average mass value
$m_R$~\cite{PDG2012} and the range is chosen according to the
typical mass resolution obtained at \lhcb.

Particle identification (PID) requirements are applied in addition to the
selection described so far.
The charged pion tracks from the \KS decay and the charged tracks from the \B
decay are all required to be inconsistent with the muon track hypothesis.
The logarithm of the likelihood ratio between the kaon and pion hypotheses
(\dllkpi), mostly based on information from the RICH detectors~\cite{LHCb-DP-2012-003},
is used to discriminate between pion and kaon candidates from the \B decay.
Pion (kaon) candidates are required to satisfy $\dllkpi<0$ ($\dllkpi>5$).
These are also required to be inconsistent with the proton hypothesis, in
order to remove the possible contributions from charmless \bquark-baryon
decays.
Pion (kaon) candidates are required to satisfy $\dllppi<10$ ($\dllpk<10$).

\section{Fit model}
\label{sec:FitModel}

A simultaneous unbinned extended maximum likelihood fit to the \B-candidate
invariant mass distributions of all decay channels is performed for each of
the two BDT optimisations.
In each simultaneous fit four types of components contribute, namely signal
decays, cross-feed backgrounds, partially-reconstructed backgrounds, and
combinatorial background.

Contributions from \BdstoKshhp decays with correct identification of the
final state particles are modelled with sums of two Crystal Ball (CB)
functions~\cite{Skwarnicki:1986xj} that share common values for the
peak position and width but have independent power law tails on opposite
sides of the peak.
The \Bd and \Bs masses (peak positions of the double-CB functions) are free
in the fit.  Four parameters related to the widths of the double-CB function
are also free parameters of the fit: the common width of the \BdtoKsPiPi
and \BstoKsPiPi signals; the relative widths of \KsKPi and \KsKK to
\KsPiPi, which are the same for \Bd and \Bs decay modes; the ratio of \LL
over \DD widths, which is the same for all decay modes.
These assumptions are made necessary by the otherwise poor determination of
the width of the suppressed mode of each spectrum. 
The other parameters of the CB components are obtained by a simultaneous
fit to simulated samples, constraining the fraction of events in the
two CB components and the ratio of their tail parameters to be the same for
all double-CB contributions.

Each selected candidate belongs uniquely to one reconstructed final state,
by definition of the particle identification criteria.
However, misidentified decays yield some cross-feed in the samples and are
modelled empirically by single CB functions using simulated events.
Only contributions from the decays \BdtoKsPiPi and \BdtoKsKK reconstructed
and selected as \KsKPi, or the decays \BstoKsKPi and \BdtoKsKPi
reconstructed and selected as either \KsKK or \KsPiPi are considered.
Other potential contributions are neglected.
The relative yield of each misidentified decay is constrained with respect
to the yield of the corresponding correctly identified decay.
The constraints are implemented using Gaussian priors included in the
likelihood. The mean values are obtained from the ratio of selection
efficiencies and the resolutions include uncertainties originating from 
the finite size of the simulated events samples and the systematic 
uncertainties related to the determination of the PID efficiencies.

Partially reconstructed charmed transitions such as $\Bm \to \Dz\pim(\Km)$
followed by $\Dz \to \KsPiPi$, with a pion not reconstructed, are
expected to dominate the background contribution in the lower invariant
mass region.
Charmless backgrounds such as from \BdtoetapKs, \BstoKstKstbartoKsPizKPi
and \ButoKsPiPiPi decays are also expected to contribute with lower rates.
These decays are modelled by means of generalised ARGUS
functions~\cite{Albrecht:1990cs} convolved with a Gaussian resolution
function.
Their parameters are determined from simulated samples. In order to
reduce the number of components in the fit, only generic contributions
for hadronic charmed and charmless decays are considered in each final
state, however \Bd and \Bs contributions are explicitly included.
Radiative decays and those from \BdtoetapKs are considered separately and
included only in the \KsPiPi final state.
The normalisation of all such contributions is constrained with Gaussian
priors using the ratio of efficiencies from the simulation and the ratio of
branching fractions from world averages \cite{PDG2012}. Relative
uncertainties on these ratios of 100\%, 20\% and 10\% are considered for
charmless, charmed, and radiative and \BdtoetapKs decays, respectively.

The combinatorial background is modelled by an exponential function, where
the slope parameter is fitted for each of the two \KS reconstruction
categories. 
The combinatorial backgrounds to the three final states \BdstoKsPiPi,
\BdstoKsKPi and \BdstoKsKK are assumed to have identical slopes. This
assumption as well as the choice of the exponential model are sources of
systematic uncertainties. 

The fit results for the two BDT optimisations are displayed in
\figstwo{fitLoose}{fitTight}.
\tab{FitResults} summarises the fitted yields of each decay mode
for the optimisation used to determine the branching fractions.
In the tight BDT optimisation the combinatorial background is negligible in
the high invariant-mass region for the \KsPiPi and \KsKK final states,
leading to a small systematic uncertainty related to the assumptions used
to fit this component.
An unambiguous first observation of \BstoKsKPi decays and a clear
confirmation of the \babar observation \cite{delAmoSanchez:2010ur} of
\BdtoKsKPi decays are obtained.
Significant yields for the \BstoKsPiPi decays are observed above negligible
background with the tight optimisation of the selection.
The likelihood profiles are shown in \fig{lhood} for \DD and \LL \KS
samples separately. 
The \BstoKsPiPi decays are observed with a combined statistical
significance of $6.2\,\sigma$, which becomes $5.9\,\sigma$ including fit
model systematic uncertainties.
The statistical significance of the \BstoKsKK signal is at the level of
$2.1\,\sigma$ combining \DD and \LL \KS reconstruction categories. 

\begin{table}[t]
\caption{Yields obtained from the simultaneous fit corresponding to the
chosen optimisation of the selection for each mode, where the uncertainties
are statistical only.
The average selection efficiencies are also given for each decay mode,
where the uncertainties are due to the limited simulation sample size.
}
\label{tab : FitResults}
\begin{center}
\begin{tabular}{l | c | >{\hfill} p{0.7cm}@{$\,\pm\,$}p{0.5cm}  >{\hfill} p{1.25cm}@{$\,\pm\,$}p{1.25cm} | >{\hfill} p{0.7cm}@{$\,\pm\,$}p{0.5cm}  >{\hfill} p{1.25cm}@{$\,\pm\,$}p{1.25cm} }
            &       & \multicolumn{4}{c|}{\DD}                                         & \multicolumn{4}{c}{\LL}                                         \\
Mode        & BDT   & \multicolumn{2}{c}{Yield} & \multicolumn{2}{c|}{Efficiency (\%)} & \multicolumn{2}{c}{Yield} & \multicolumn{2}{c}{Efficiency (\%)} \\
\hline
\BdtoKsPiPi & Loose & $845$ & $38$              & $0.0336$ & $0.0010$                  & $360$ & $21$              & $0.0117$ & $0.0009$                 \\
\BdtoKsKK   & Loose & $256$ & $20$              & $0.0278$ & $0.0008$                  & $175$ & $15$              & $0.0092$ & $0.0016$                 \\
\BstoKsKPi  & Loose & $283$ & $24$              & $0.0316$ & $0.0007$                  & $152$ & $15$              & $0.0103$ & $0.0008$                 \\
\BdtoKsKPi  & Tight & $92$  & $15$              & $0.0283$ & $0.0009$                  & $52$  & $11$              & $0.0133$ & $0.0005$                 \\
\BstoKsPiPi & Tight & $28$  & $9$               & $0.0153$ & $0.0013$                  & $25$  & $6$               & $0.0109$ & $0.0006$                 \\
\BstoKsKK   & Tight & $6$   & $4$               & $0.0150$ & $0.0021$                  & $3$   & $3$               & $0.0076$ & $0.0016$                 \\
\end{tabular}
\end{center}
\end{table} 

\begin{figure}[!htbp]
  \begin{center}
  \ifthenelse{\boolean{pdflatex}}{
    \includegraphics*[width=0.49\textwidth]{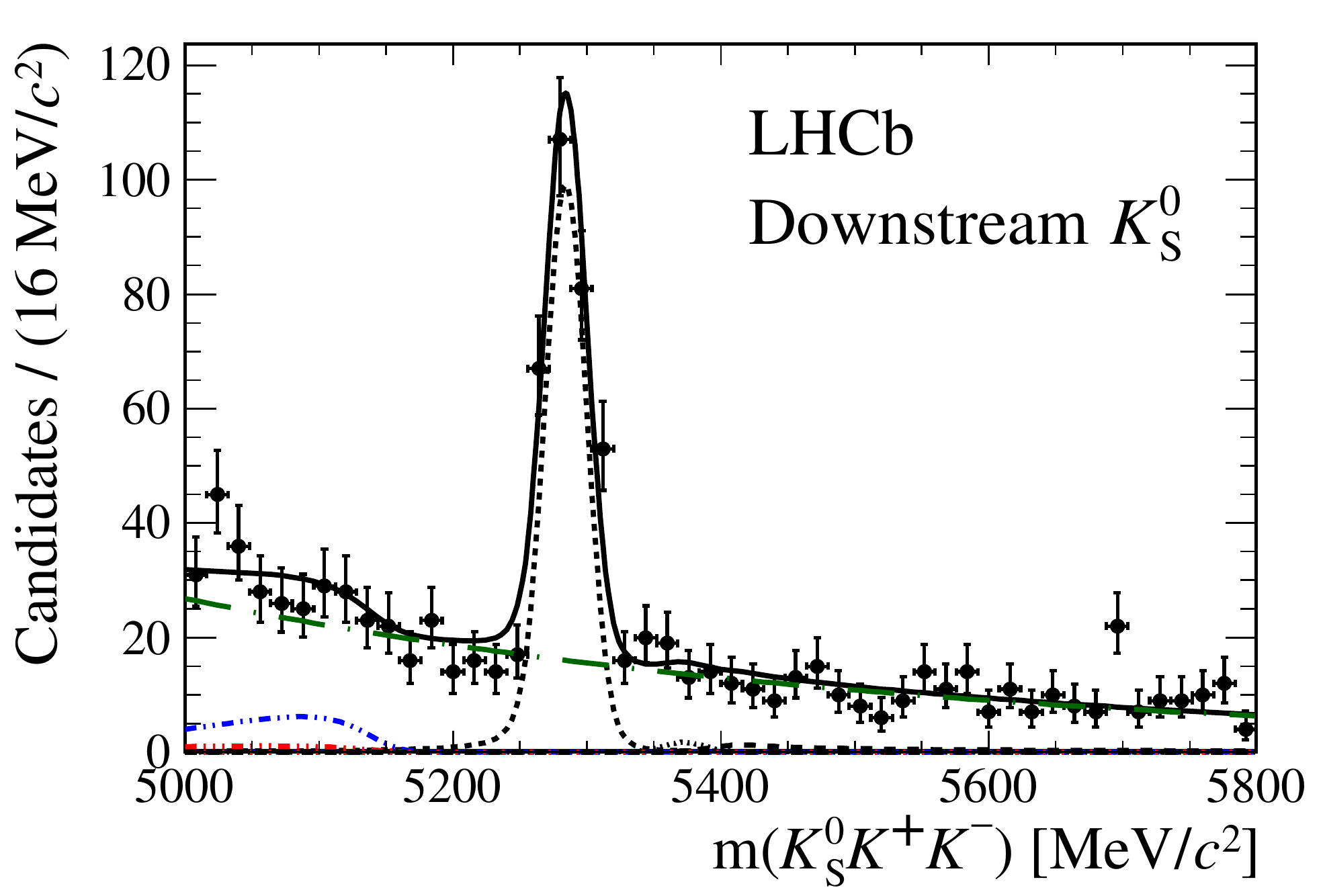}
    \includegraphics*[width=0.49\textwidth]{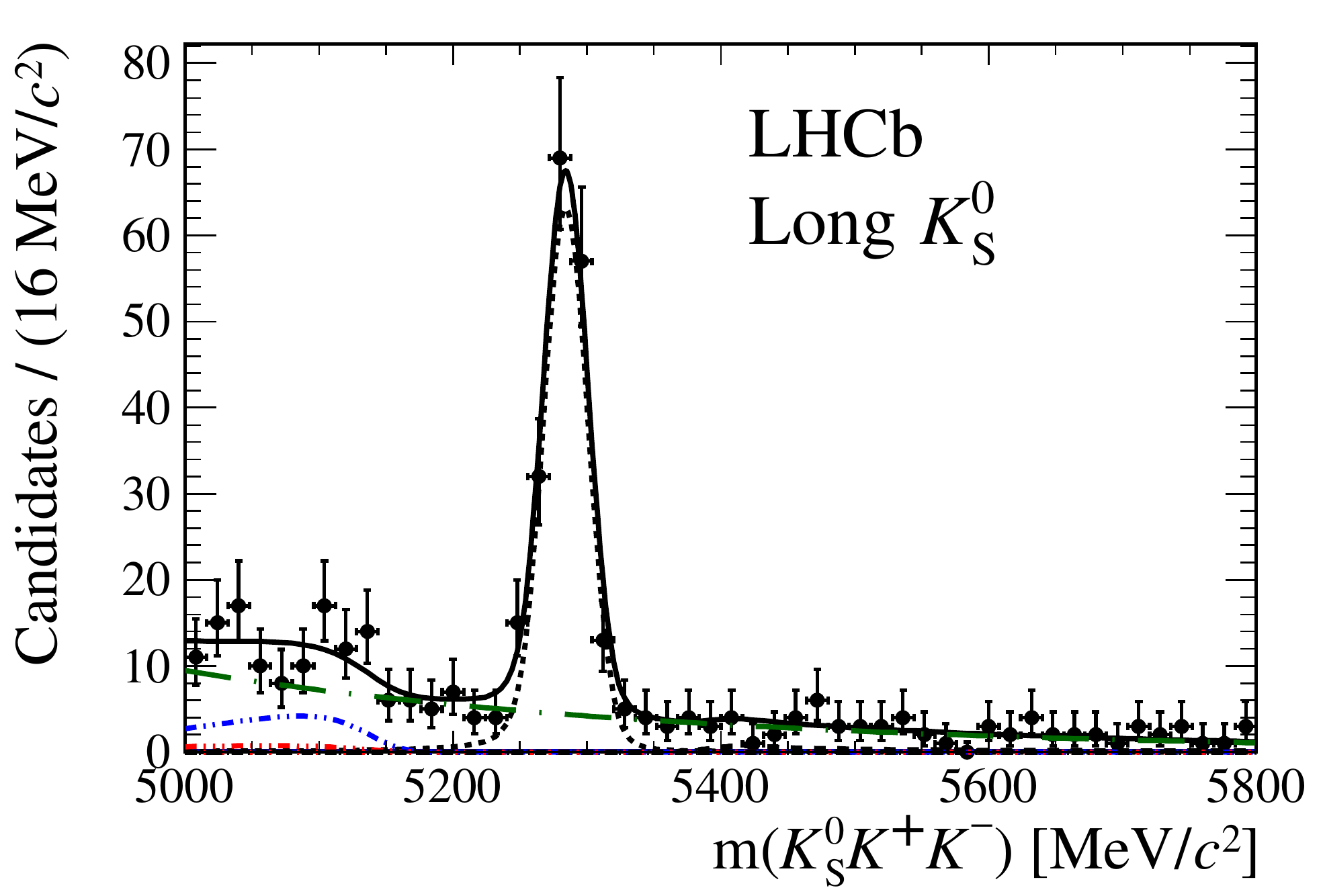}\\
    \vspace{2ex}
    \includegraphics*[width=0.49\textwidth]{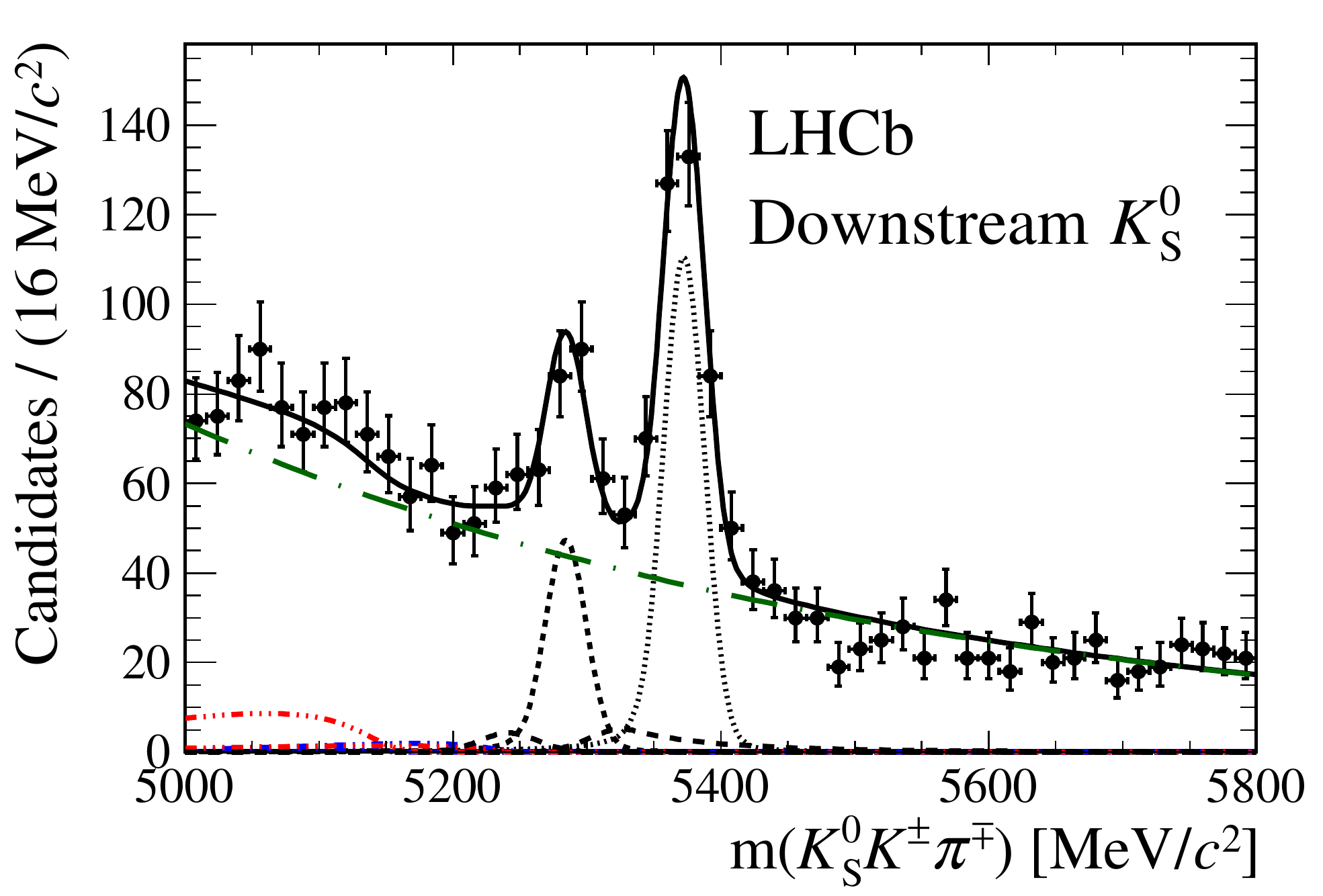}
    \includegraphics*[width=0.49\textwidth]{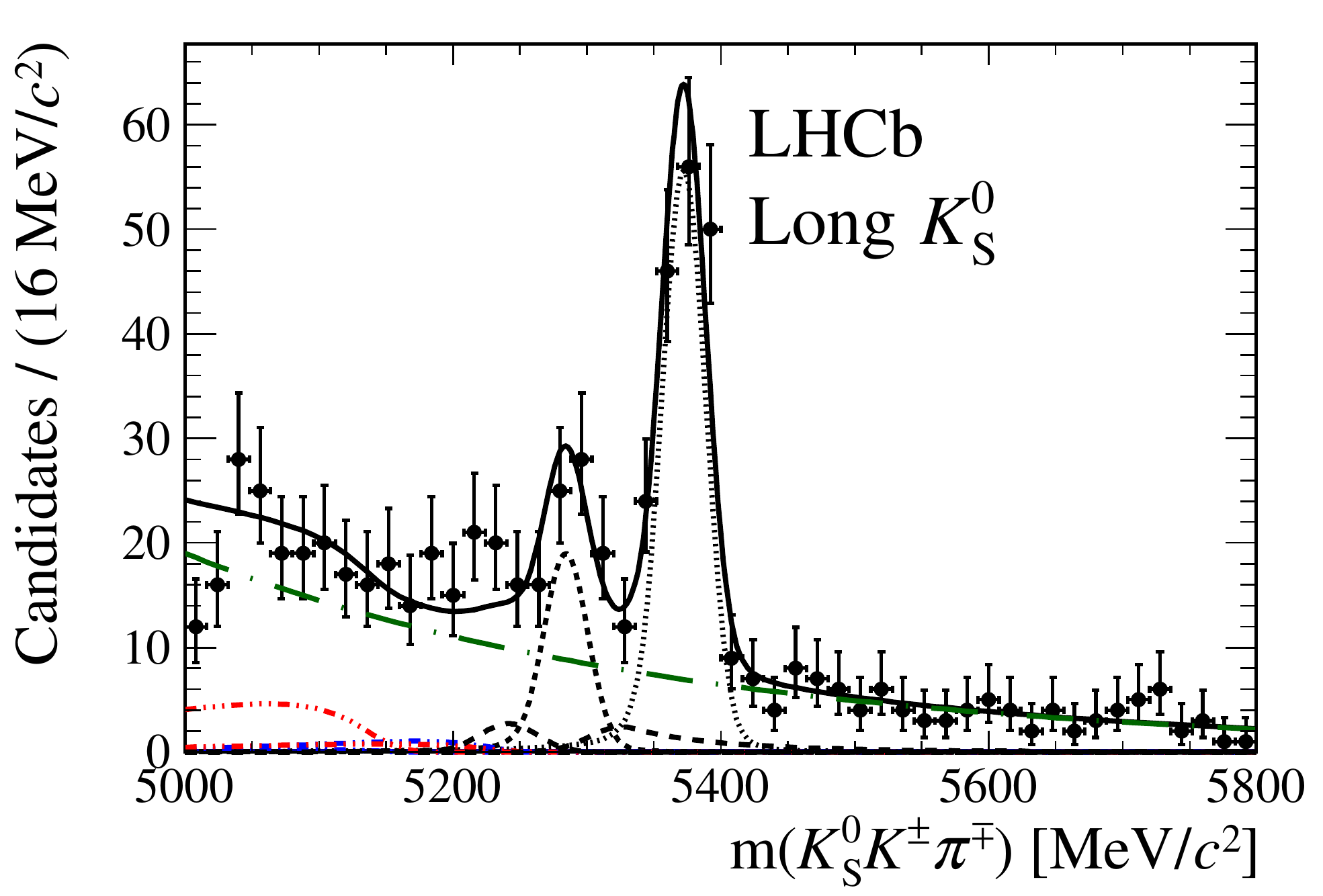}\\
    \vspace{2ex}
    \includegraphics*[width=0.49\textwidth]{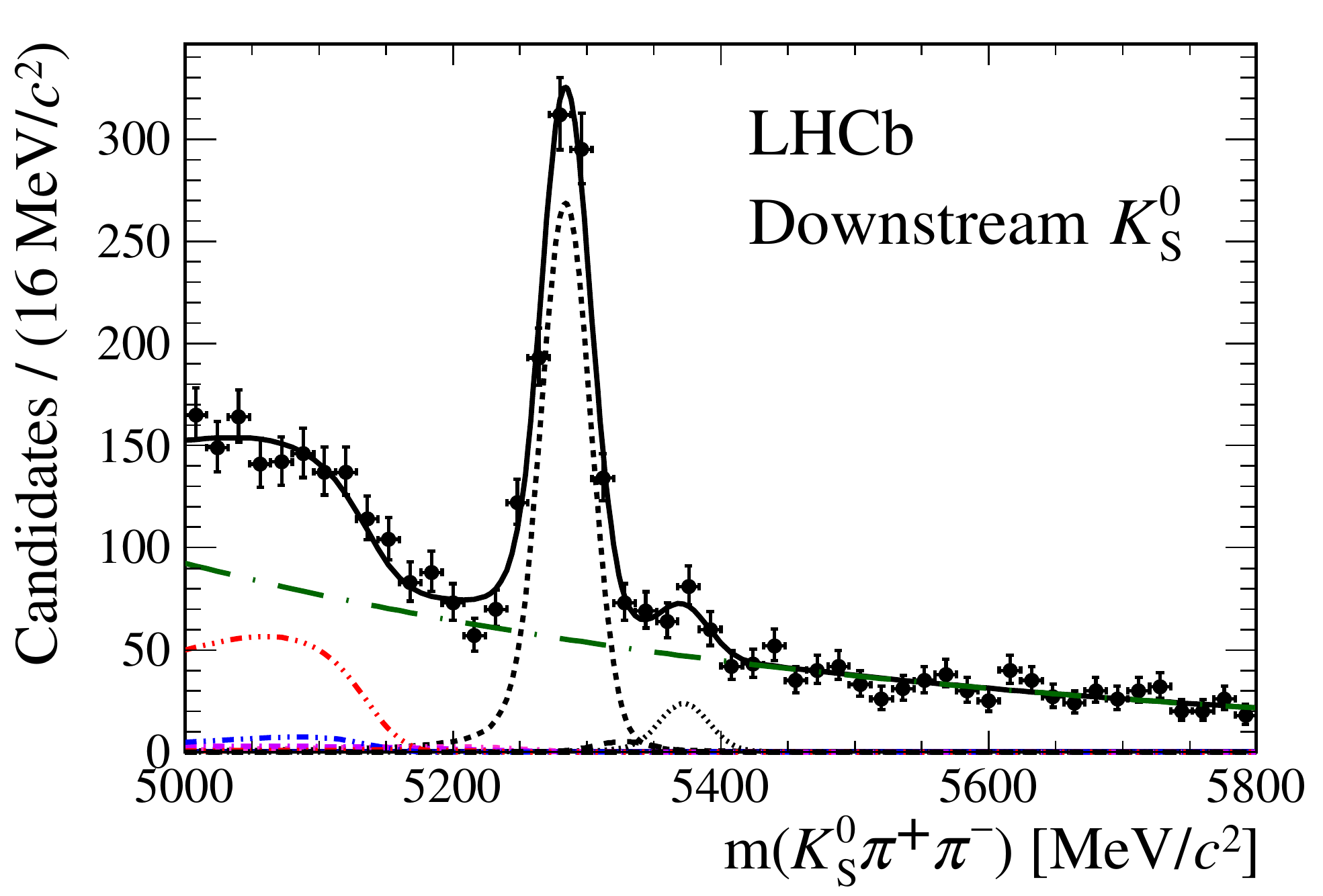}
    \includegraphics*[width=0.49\textwidth]{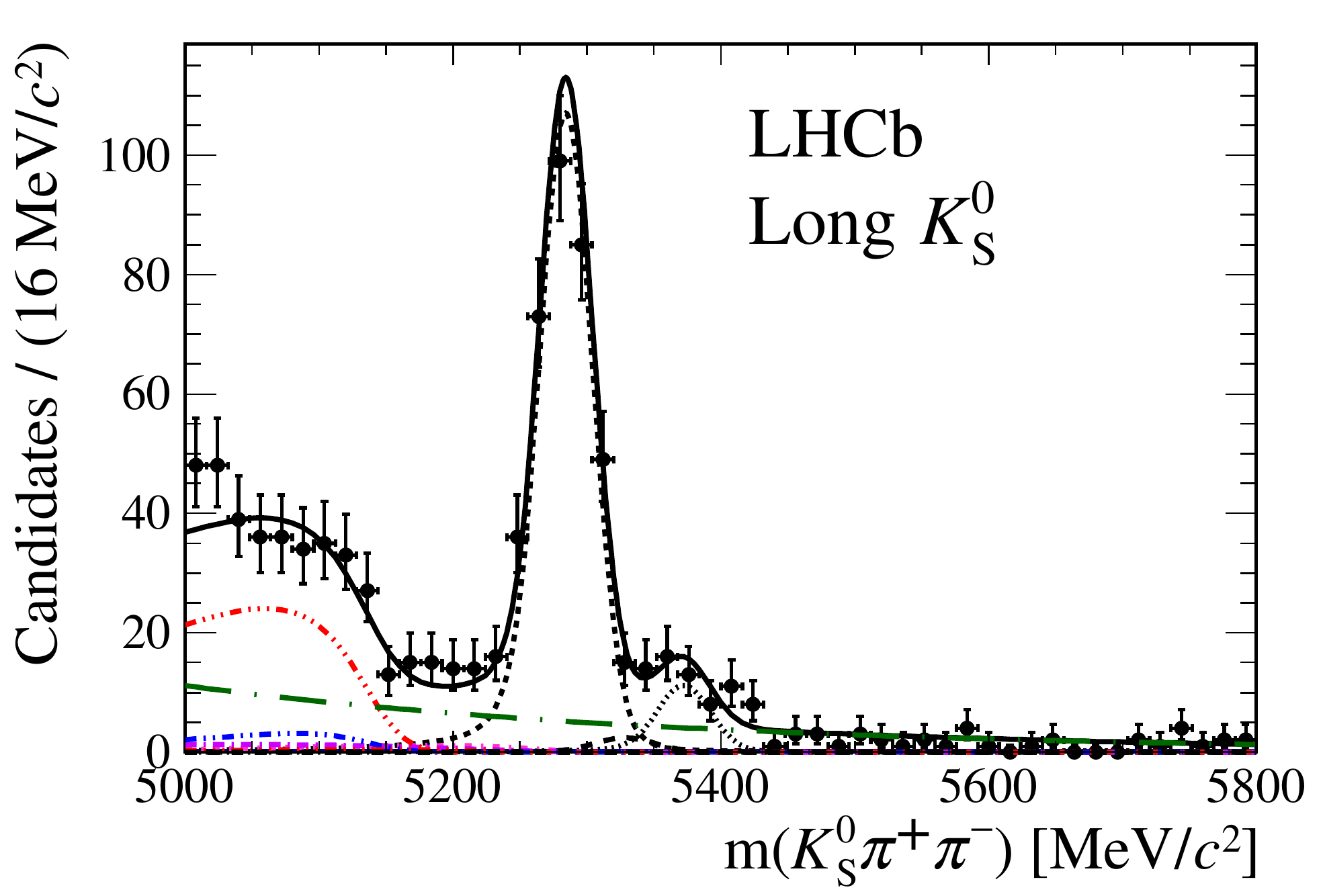}
  }{
    \includegraphics*[width=0.49\textwidth]{figs/cBmass_KSKK_DDdata2011-bdtBR-All-bdtBRfixSlopes-paper.eps}
    \includegraphics*[width=0.49\textwidth]{figs/cBmass_KSKK_LLdata2011-bdtBR-All-bdtBRfixSlopes-paper.eps}\\
    \vspace{2ex}
    \includegraphics*[width=0.49\textwidth]{figs/cBmass_KSKpi_DDdata2011-bdtBR-All-bdtBRfixSlopes-paper.eps}
    \includegraphics*[width=0.49\textwidth]{figs/cBmass_KSKpi_LLdata2011-bdtBR-All-bdtBRfixSlopes-paper.eps}\\
    \vspace{2ex}
    \includegraphics*[width=0.49\textwidth]{figs/cBmass_KSpipi_DDdata2011-bdtBR-All-bdtBRfixSlopes-paper.eps}
    \includegraphics*[width=0.49\textwidth]{figs/cBmass_KSpipi_LLdata2011-bdtBR-All-bdtBRfixSlopes-paper.eps}
   }
  \end{center}
  \caption{\small
  Invariant mass distributions of (top) \KsKK, (middle) \KsKPi, and (bottom)
  \KsPiPi candidate events, with the loose selection for (left) \DD and
  (right) \LL \KS reconstruction categories.
  In each plot, data are the black points with error bars and the total fit
  model is overlaid (solid black line).
  The \Bd (\Bs) signal components are the black short-dashed (dotted) lines,
  while fully reconstructed misidentified decays are the black dashed lines
  close to the \Bd and \Bs peaks.
  The partially reconstructed contributions from \B to open charm decays,
  charmless hadronic decays, \BdtoetapKs and charmless radiative decays are
  the red dash triple-dotted, the blue dash double-dotted, the violet dash
  single-dotted, and the pink short-dash single-dotted lines, respectively.
  The combinatorial background contribution is the green long-dash dotted
  line.
  }
  \label{fig : fitLoose}
\end{figure}

\begin{figure}[!htbp]
  \begin{center}
  \ifthenelse{\boolean{pdflatex}}{
    \includegraphics*[width=0.49\textwidth]{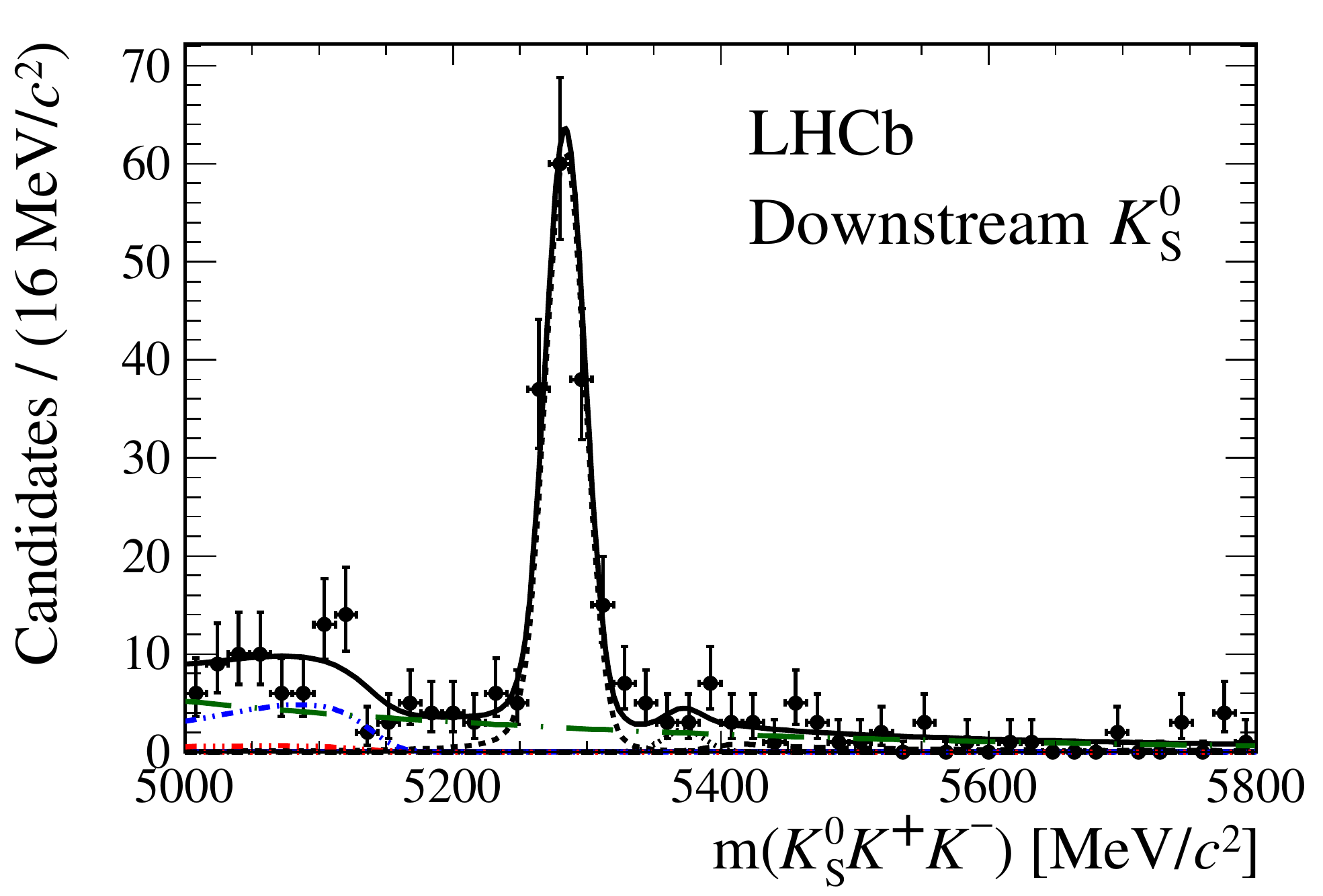}
    \includegraphics*[width=0.49\textwidth]{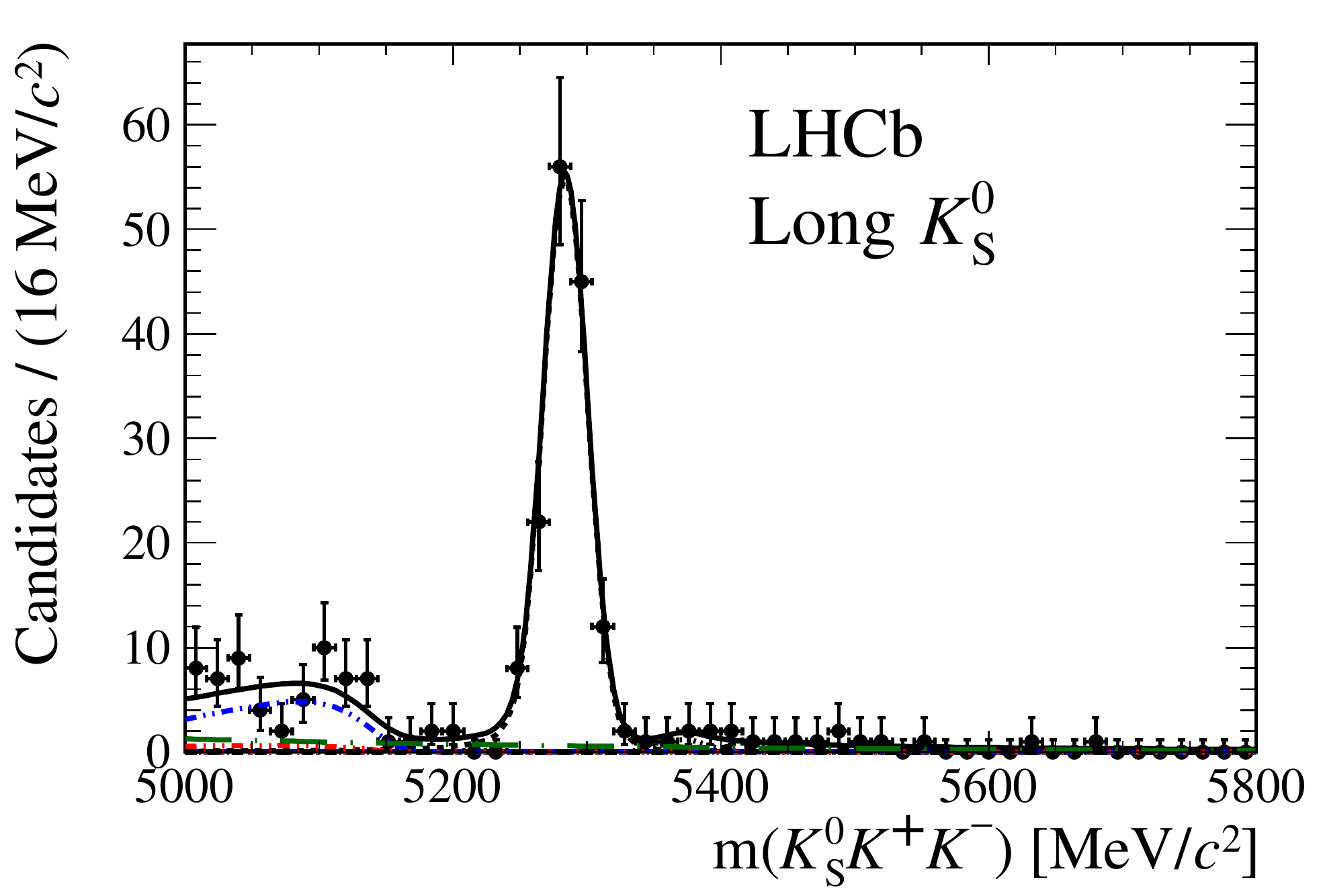}\\
    \vspace{2ex}
    \includegraphics*[width=0.49\textwidth]{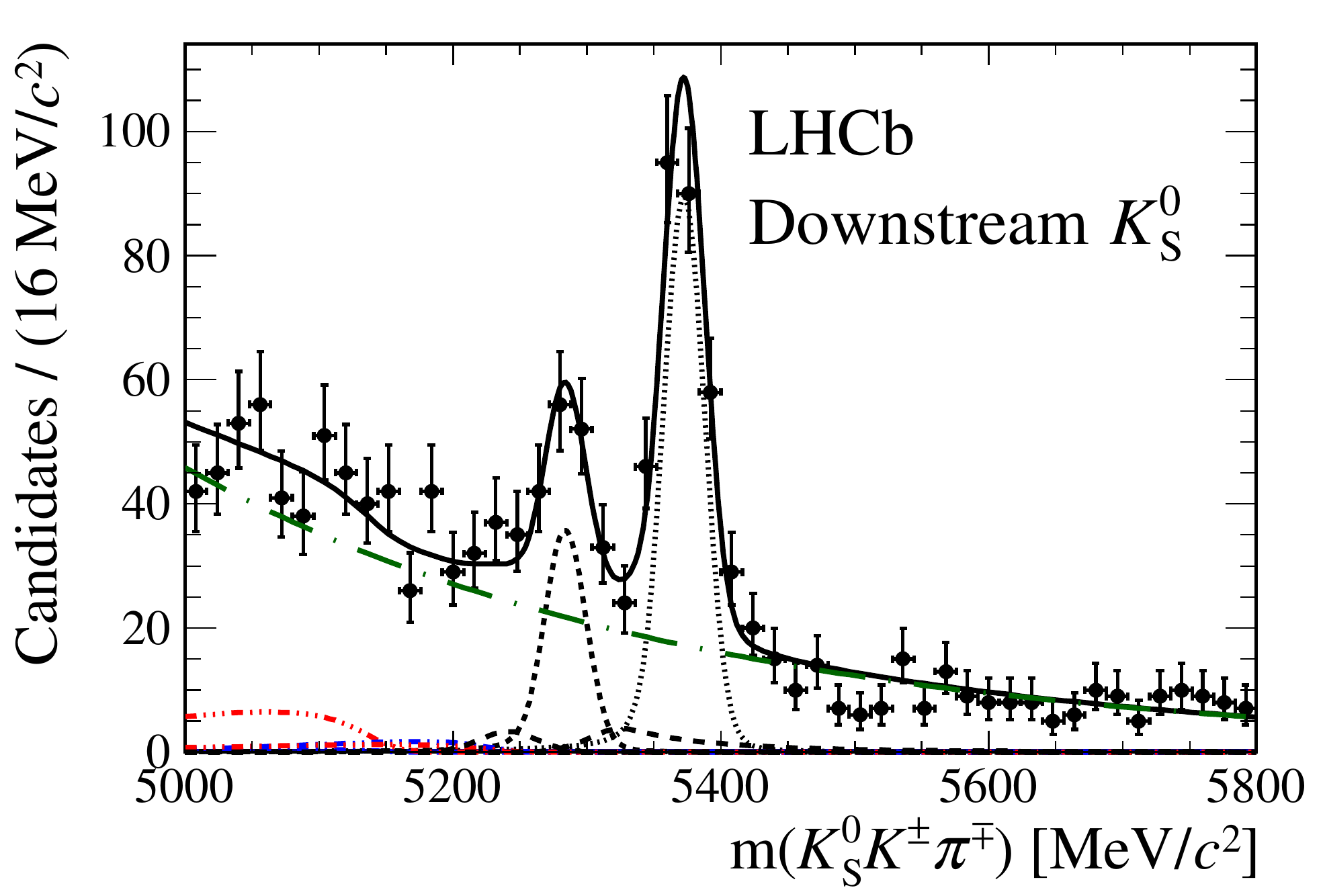}
    \includegraphics*[width=0.49\textwidth]{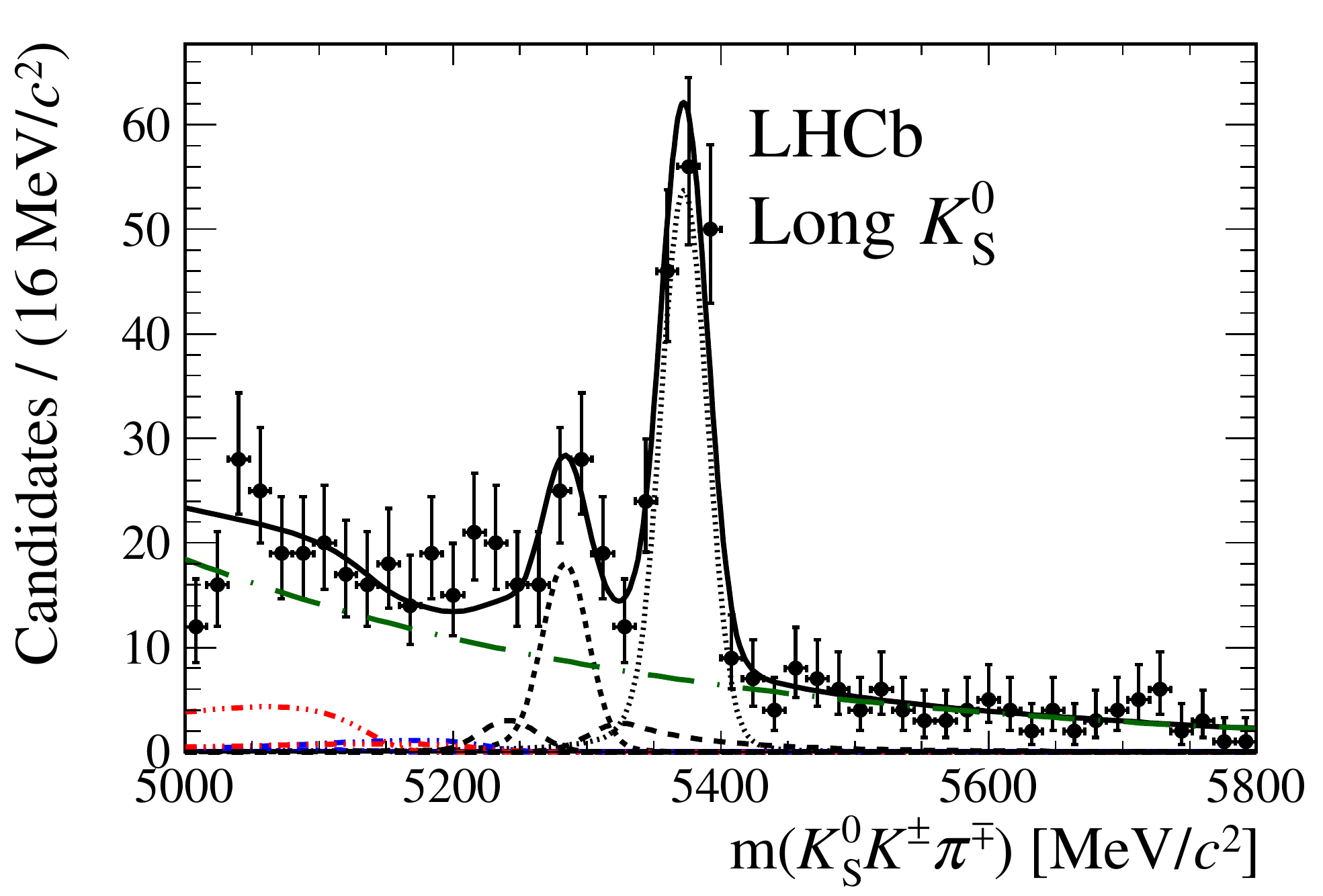}\\
    \vspace{2ex}
    \includegraphics*[width=0.49\textwidth]{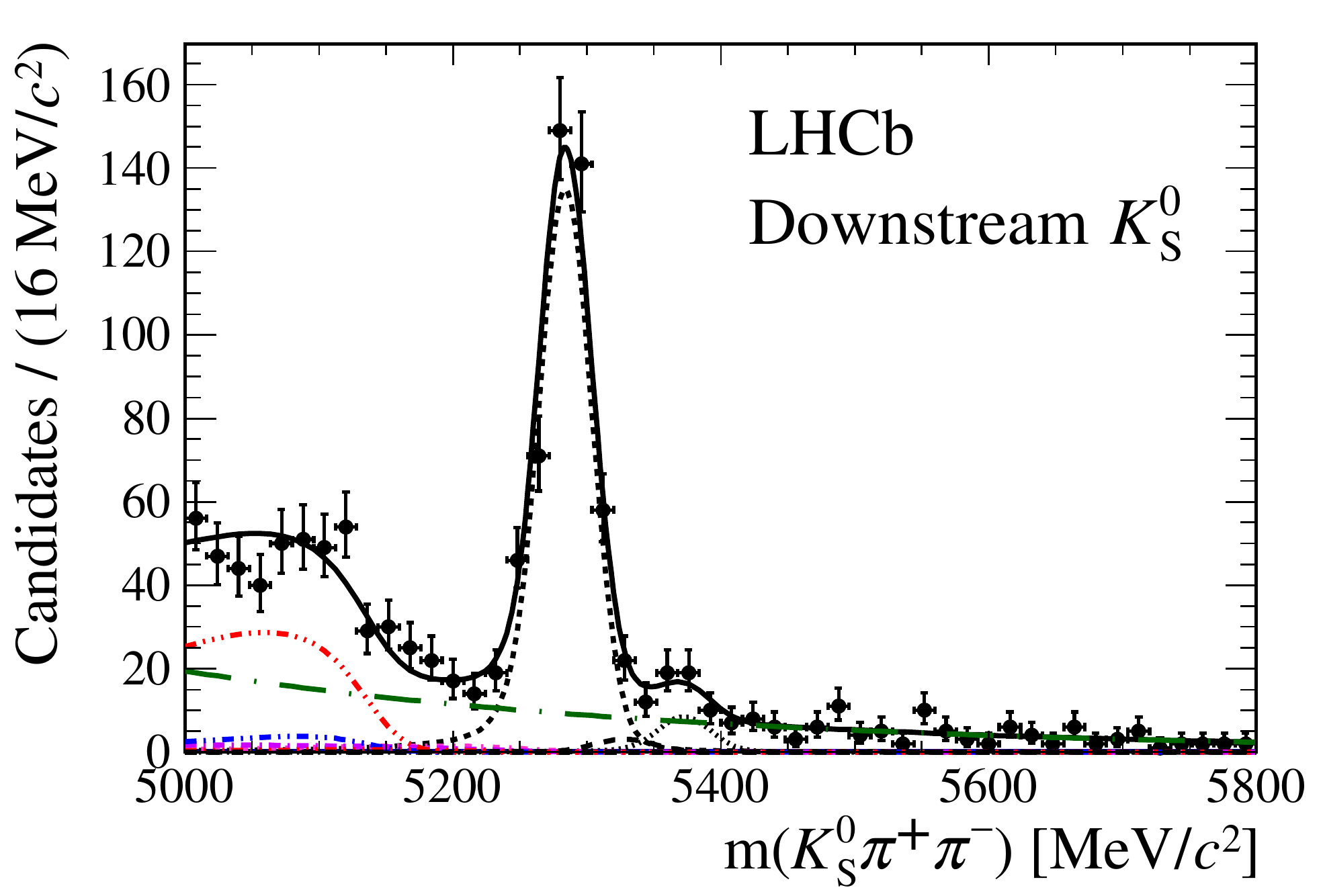}
    \includegraphics*[width=0.49\textwidth]{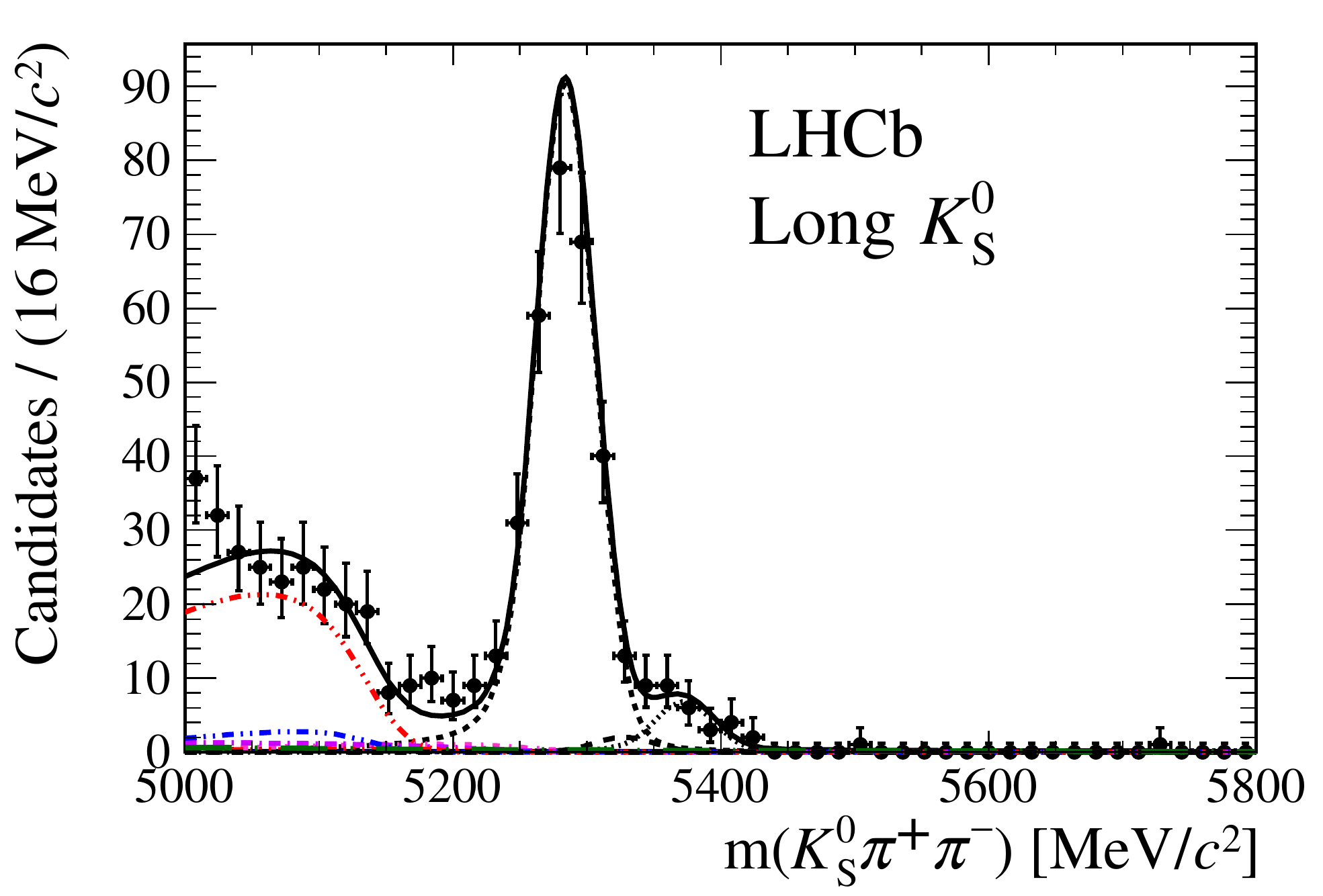}
  }{
    \includegraphics*[width=0.49\textwidth]{figs/cBmass_KSKK_DDdata2011-bdtOBS-All-bdtOBSfixSlopes-paper.eps}
    \includegraphics*[width=0.49\textwidth]{figs/cBmass_KSKK_LLdata2011-bdtOBS-All-bdtOBSfixSlopes-paper.eps}\\
    \vspace{2ex}
    \includegraphics*[width=0.49\textwidth]{figs/cBmass_KSKpi_DDdata2011-bdtOBS-All-bdtOBSfixSlopes-paper.eps}
    \includegraphics*[width=0.49\textwidth]{figs/cBmass_KSKpi_LLdata2011-bdtOBS-All-bdtOBSfixSlopes-paper.eps}\\
    \vspace{2ex}
    \includegraphics*[width=0.49\textwidth]{figs/cBmass_KSpipi_DDdata2011-bdtOBS-All-bdtOBSfixSlopes-paper.eps}
    \includegraphics*[width=0.49\textwidth]{figs/cBmass_KSpipi_LLdata2011-bdtOBS-All-bdtOBSfixSlopes-paper.eps}
   }
  \end{center}
  \caption{\small
  Invariant mass distributions of (top) \KsKK, (middle) \KsKPi, and (bottom)
  \KsPiPi candidate events, with the tight selection for (left) \DD and
  (right) \LL \KS reconstruction categories.
  In each plot, data are the black points with error bars and the total fit
  model is overlaid (solid black line).
  The \Bd (\Bs) signal components are the black short-dashed (dotted) lines,
  while fully reconstructed misidentified decays are the black dashed lines
  close to the \Bd and \Bs peaks.
  The partially reconstructed contributions from \B to open charm decays,
  charmless hadronic decays, \BdtoetapKs and charmless radiative decays are
  the red dash triple-dotted, the blue dash double-dotted, the violet dash
  single-dotted, and the pink short-dash single-dotted lines, respectively.
  The combinatorial background contribution is the green long-dash dotted
  line.
  }
  \label{fig : fitTight}
\end{figure}

\begin{figure}[!htbp]
  \begin{center}
  \ifthenelse{\boolean{pdflatex}}{
    \includegraphics*[width=0.49\textwidth]{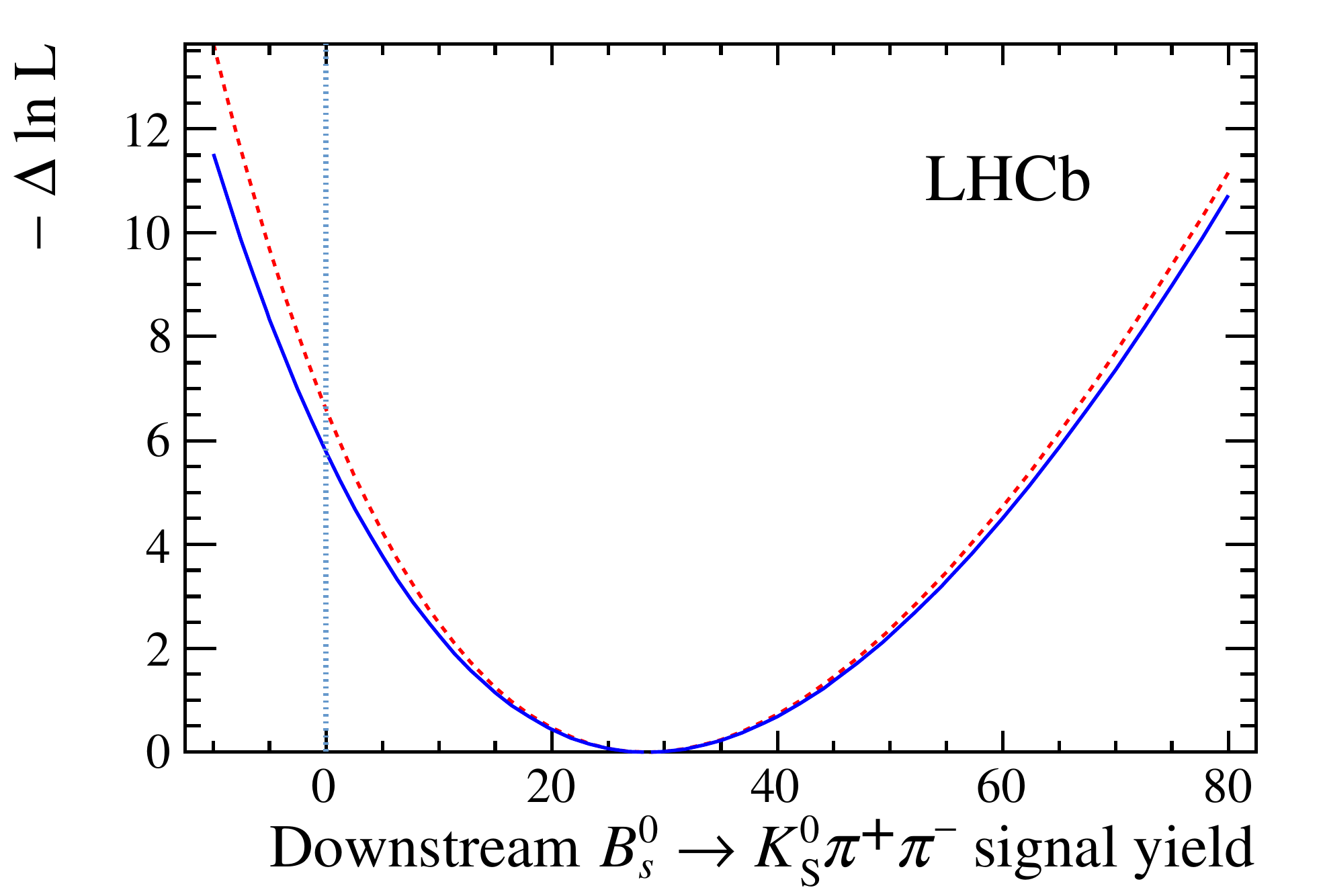}
    \includegraphics*[width=0.49\textwidth]{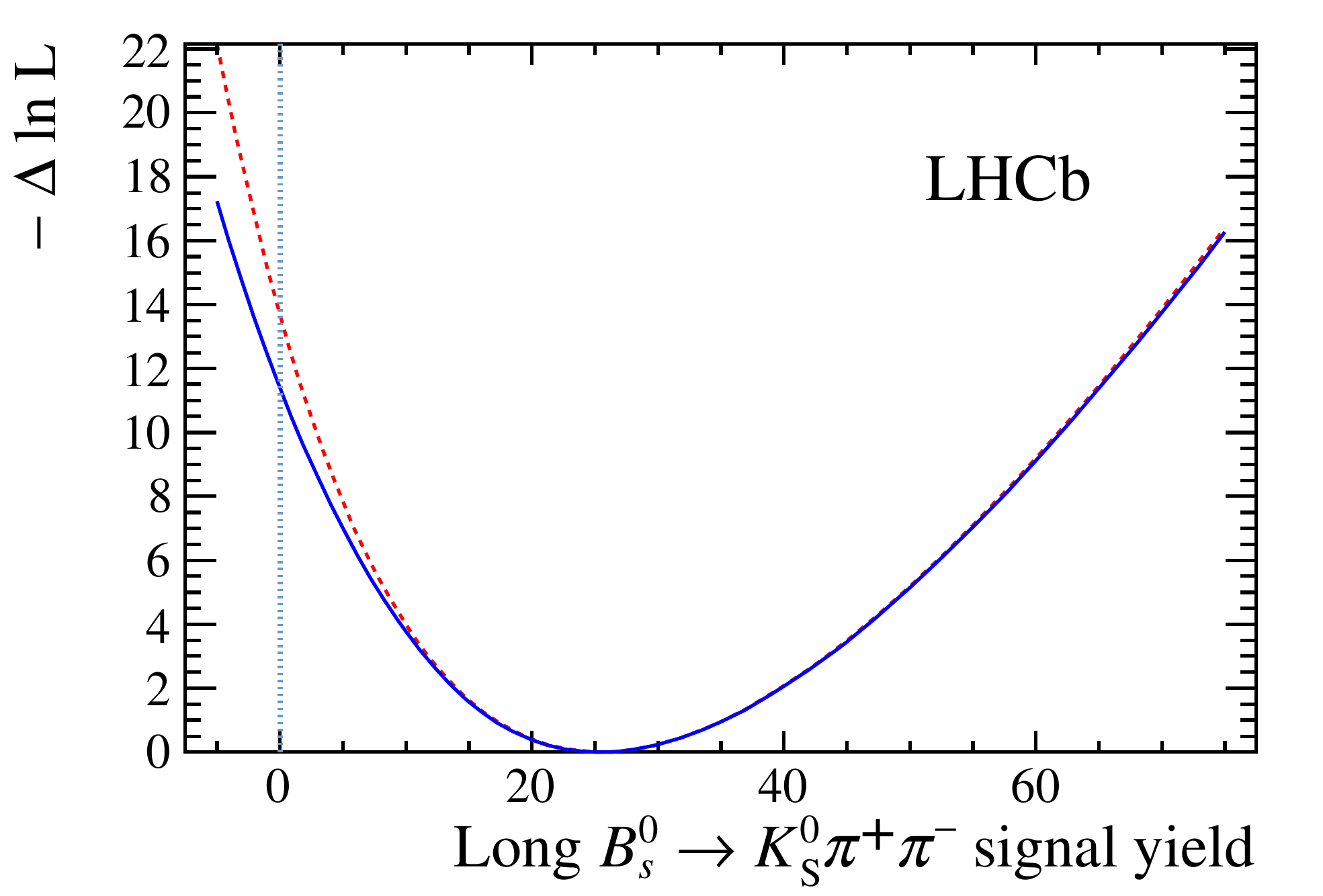}
    }{
    \includegraphics*[width=0.49\textwidth]{figs/scan-with-syst_DD.eps}
    \includegraphics*[width=0.49\textwidth]{figs/scan-with-syst_LL.eps}
    }
  \end{center}
 \caption{\small Likelihood profiles of the \BstoKsPiPi signal yield for
 the (left) \DD and (right) \LL \KS samples. The dashed red line is the
 statistical-only profile, while the solid blue line also includes the
 fit model systematic uncertainties. The significance of the \DD and \LL signals are
 $3.4\,\sigma$ and $4.8\,\sigma$, respectively, including systematic
 uncertainties. Combining \DD and \LL \KS samples, an observation with
 $5.9\,\sigma$, including systematic uncertainties, is obtained.}
  \label{fig : lhood}
\end{figure}

\section{Determination of the efficiencies} 
\label{sec:Efficiencies}

The measurements of the branching fractions of the \BdstoKshhp decays
relative to the well established  \BdtoKsPiPi decay mode proceed according
to
\begin{eqnarray}      
\frac{\BF(\BdstoKshhp)}{\BF(\BdtoKsPiPi)} & = &
\frac{\eps^{\rm sel}_{\BdtoKsPiPi}}{\eps^{\rm sel}_{\BdstoKshhp}}
\frac{N_{\BdstoKshhp} }{ N_{\BdtoKsPiPi}} \frac{f_d}{f_{d,s}}\,,
\end{eqnarray}  
where $\eps^{\rm sel}$ is the selection efficiency (which includes
acceptance, reconstruction, selection, trigger and particle identification
components),
$N$ is the fitted signal yield, and $f_d$ and $f_s$ are the hadronisation
fractions of a \bquark quark into a \Bz and \Bs meson, respectively.
The ratio \fsfdinline has been accurately determined by the \lhcb
experiment from hadronic and semileptonic measurements
$\fsfdinline = 0.256 \pm 0.020$~\cite{LHCb-PAPER-2012-037}. 

Three-body decays are composed of several quasi-two-body decays and
non-resonant contributions, all of them possibly interfering.
Hence, their dynamical structure, described by the
Dalitz plot~\cite{Dalitz:1953cp}, must be accounted for to correct for
non-flat efficiencies over the phase space.
Since the dynamics of most of the modes under study are not known prior to
this analysis, efficiencies are determined for each decay mode from
simulated signal samples in bins of the ``square Dalitz plot''~\cite{Aubert:2005sk},
where the usual Dalitz-plot coordinates have been transformed into a
rectangular space.
The edges of the usual Dalitz plot are spread out in the square
Dalitz plot, which permits a more precise modelling of the efficiency
variations in the regions where they are most strongly varying and where
most of the signal events are expected.
Two complementary simulated samples have been produced, corresponding to
events generated uniformly in phase space or uniformly in the square Dalitz
plot.
The square Dalitz-plot distribution of each signal mode is determined from
the data using the \sPlot\ technique~\cite{Pivk:2004ty}.
The binning is chosen such that each bin is populated by approximately the
same number of signal events.
The average efficiency for each decay mode is calculated as the weighted
harmonic mean over the bins.
The average weighted selection efficiencies are summarised in
\tab{FitResults} and depend on the final state, the \KS reconstruction
category, and the choice of the BDT optimisation.
Their relative uncertainties due to the finite size of the simulated event
samples vary from 3\% to 17\%, reflecting the different dynamical
structures of the decay modes.  

The particle identification and misidentification efficiencies are determined from 
simulated signal events on an event-by-event basis by adjusting the DLL
distributions measured from calibration events to match the kinematical properties
of the tracks in the decay of interest. The reweighting is performed in bins of \ptot 
and \pt, accounting for kinematic correlations between the tracks.  Calibration tracks
are taken from $\Dstarp \to \Dz \pip_s$ decays where the \Dz decays to the
Cabibbo-favoured $\Km\pip$ final state.  The charge of the soft pion $\pip_s$ hence
provides the kaon or pion identity of the tracks. 
The dependence of the PID efficiency over the Dalitz plot is included in
the procedure described above. 
This calibration is performed using samples from the same data
taking period, accounting for the variation in the performance of the RICH detectors
over time.

\section{Systematic uncertainties}
\label{sec : systematics}

Most of the systematic uncertainties are eliminated or greatly reduced by
normalising the branching fraction measurements with respect to the
\BdtoKsPiPi mode.
The remaining sources of systematic effects and the methods used to
estimate the corresponding uncertainties are described in this section.
In addition to the systematic effects related to the measurements performed
in this analysis, there is that associated with the measured value of
\fsfdinline.
A summary of the contributions, expressed as relative uncertainties, is
given in \tab{syst_BR_DDLL}.

\subsection{Fit model}
\label{sec : fitmodel syst}

The fit model relies on a number of assumptions, both in the values of
parameters being taken from simulation and in the choice of the functional
forms describing the various contributions.

The uncertainties linked to the parameters fixed to values determined from
simulated events are obtained by repeating the fit while the fixed
parameters are varied according to their uncertainties using pseudo-experiments.
For example, the five fixed parameters of the CB functions describing the
signals, as well as the ratio of resolutions with respect to \BdtoKsPiPi
decays, are varied according to their correlation matrix determined from
simulated events. The nominal fit is then performed on this sample of
pseudo-experiments and the distribution of the difference between the yield
determined in each of these fits and that of the nominal fit is fitted with
a Gaussian function. The systematic uncertainty associated with the choice
of the value of each signal parameter from simulated events is then
assigned as the linear sum of the absolute value of the mean of the
Gaussian and its resolution.  An identical procedure is employed to obtain
the systematic uncertainties related to the fixed parameters of the \argus
functions describing the partially reconstructed backgrounds and the CB
functions used for the cross-feeds.

The uncertainties related to the choice of the models used in the nominal
fit are evaluated for the signal and combinatorial background models only.
Both the partially reconstructed background and the cross-feed shapes
suffer from a large statistical uncertainty from the simulated event
samples and therefore the uncertainty related to the fixed parameters also
covers any sensible variation of the shape.
The \Bs decay modes that are studied lie near large \Bd contributions for
the \KsPiPi and \KsKK spectra. The impact of the modelling of  the right
hand side of the \Bd mass distribution is addressed by removing the second
CB function, used as an alternative model.

For the combinatorial background, a unique slope parameter governs the
shape of each \KS reconstruction category (\LL or \DD). Two alternative
models are considered: allowing independent slopes for each of the six
spectra (testing the assumption of a universal slope) and using a linear
model in place of the exponential (testing the functional form of the 
combinatorial shape). Pseudo-experiments are again used to estimate the
effect of these alternative models; in the former case, the value and uncertainties
to be considered for the six slopes are determined from a fit to the data. 
The dataset is generated according to the substitute model and the fit is performed
to the generated sample using the nominal model. The value of the uncertainty 
is again estimated as the linear sum of the absolute value of the resulting bias 
and its resolution. The total fit model systematic uncertainty is given by the sum in
quadrature of all the contributions and is mostly dominated by the
combinatorial background model uncertainty.  

\subsection{Selection and trigger efficiencies}
\label{sec : eff syst}

The accuracy of the efficiency determination is limited in most cases by
the finite size of the samples of simulated signal events, duly propagated 
as a systematic uncertainty.  In addition, the effect related to the
choice of binning for the square Dalitz plot is estimated from the 
spread of the average efficiencies determined from several alternative binning
schemes.
Good agreement between data and the simulation is obtained, hence no further
systematic uncertainty is assigned.

Systematic uncertainties related to the hardware stage trigger have been 
studied.  A data control sample of $\Dstarp \to \Dz (\to \Km\pip) \pip_s$ decays is
used to quantify differences between pions and kaons, separated by positive
and negative hadron charges, as a function of \pt~\cite{LHCb-DP-2012-004}. 
Though they show an overall good agreement for the different types of tracks,
the efficiency for pions is slightly smaller than for kaons at high \pt.
Simulated events are reweighted by these data-driven calibration curves in
order to extract the hadron trigger efficiency for each mode, propagating
properly the calibration-related uncertainties.
Finally, the ageing of the calorimeters during the data taking period when
the data sample analysed was recorded induced changes in the absolute scale
of the trigger efficiencies.  While this was mostly mitigated by periodic
recalibration, relative variations occurred of order $10\%$.
Since the kinematics vary marginally from one mode to the other, a
systematic effect on the ratio of efficiencies arises.
It is fully absorbed by increasing the trigger efficiency systematic
uncertainty by $10\%$.   

\subsection{Particle identification efficiencies}
\label{sec : pid syst}

The procedure to evaluate the efficiencies of the PID selections uses calibration 
tracks that differ from the signal tracks in terms of their kinematic distributions.
While the binning procedure attempts to mitigate these differences there could
be some remaining systematic effect.  To quantify any bias due to the procedure,
simulated samples of the control modes are used in place of the data samples.
The average efficiency determined from these samples can then be compared
with the efficiency determined from simply applying the selections to the
simulated signal samples.  The differences are found to be less than $1\%$,
hence no correction is applied.
The calibration procedure is assigned a systematic uncertainty.
The observed differences in efficiencies are multiplied by the efficiency
ratio and statistical uncertainties from the finite sample sizes are added
in quadrature.

\begin{table}[t]
  \caption{Systematic uncertainties on the ratio of branching fractions for
  \DD and \LL \KS reconstruction.  All uncertainties are relative and are quoted as percentages.}
\begin{center}
\resizebox{\textwidth}{!}{
\begin{tabular}{ l | c c c c | c | c } 
\DD                                  & Fit   & Selection & Trigger  & PID   &  Total & \fsfdinline \\
\hline
 \Br{\BdtoKsKPi}  / \Br{\BdtoKsPiPi} & $ 5 $ & $  6   $  & $ 3   $  & $ 1 $ & $  8 $ & ---         \\
 \Br{\BdtoKsKK}   / \Br{\BdtoKsPiPi} & $ 1 $ & $  5   $  & $ 3   $  & $ 1 $ & $  6 $ & ---         \\
 \Br{\BstoKsPiPi} / \Br{\BdtoKsPiPi} & $ 8 $ & $ 16   $  & $ 2   $  & $ 1 $ & $ 18 $ & $   8 $     \\    
 \Br{\BstoKsKPi}  / \Br{\BdtoKsPiPi} & $ 2 $ & $  5   $  & $ 1   $  & $ 1 $ & $  6 $ & $   8 $     \\     
 \Br{\BstoKsKK}   / \Br{\BdtoKsPiPi} & $ 1 $ & $ 18   $  & $ 3   $  & $ 1 $ & $ 18 $ & $   8 $     \\
 \hline
 \LL                                 &       &           &          &       &        &             \\
\hline
 \Br{\BdtoKsKPi}  / \Br{\BdtoKsPiPi} & $ 5 $ & $ 10   $  & $ 1   $  & $ 1 $ & $ 14 $ & ---         \\
 \Br{\BdtoKsKK}   / \Br{\BdtoKsPiPi} & $ 3 $ & $ 20   $  & $ 1   $  & $ 1 $ & $ 20 $ & ---         \\
 \Br{\BstoKsPiPi} / \Br{\BdtoKsPiPi} & $ 5 $ & $ 10   $  & $ 1   $  & $ 1 $ & $ 11 $ & $   8 $     \\    
 \Br{\BstoKsKPi}  / \Br{\BdtoKsPiPi} & $ 3 $ & $ 12   $  & $ 2   $  & $ 1 $ & $ 13 $ & $   8 $     \\     
 \Br{\BstoKsKK}   / \Br{\BdtoKsPiPi} & $ 2 $ & $ 22   $  & $ 1   $  & $ 1 $ & $ 22 $ & $   8 $     \\
\end{tabular}
}
\end{center}
\label{tab : syst_BR_DDLL}
\end{table}

\section{Results and conclusion} 
\label{sec:Results}

The 2011 LHCb dataset, corresponding to an integrated luminosity of
1.0\invfb recorded at a centre-of-mass energy of 7\tev, has been analysed
to search for the decays \BdstoKshhp.
The decays \BstoKsKPi and \BstoKsPiPi are observed for the first time.
The former is unambiguous, while for the latter the significance of the
observation is $5.9$ standard deviations, including statistical and
systematic uncertainties.
The decay mode \BdtoKsKPi, previously observed by the \babar
experiment~\cite{delAmoSanchez:2010ur}, is confirmed.
The efficiency-corrected Dalitz-plot distributions of the three decay modes
\BstoKsPiPi, \BstoKsKPi, and \BdtoKsKPi are displayed in \fig{splot dalitz}. 
Some structure is evident at low $\KS\pipm$ and $\Kpm\pimp$ invariant
masses in the \BstoKsKPi decay mode, while in the \BdtoKsKPi decay the
largest structure is seen in the low $\KS\Kpm$ invariant mass region.
No significant evidence for \BstoKsKK decays is obtained. A 90\% confidence level
(CL) interval based on the CL inferences described in Ref.~\cite{Feldman:1997qc}
is hence placed on the branching fraction for this decay mode.

\begin{figure}[!htbp]
\begin{center}
\ifthenelse{\boolean{pdflatex}}{
\includegraphics[width=0.6\textwidth]{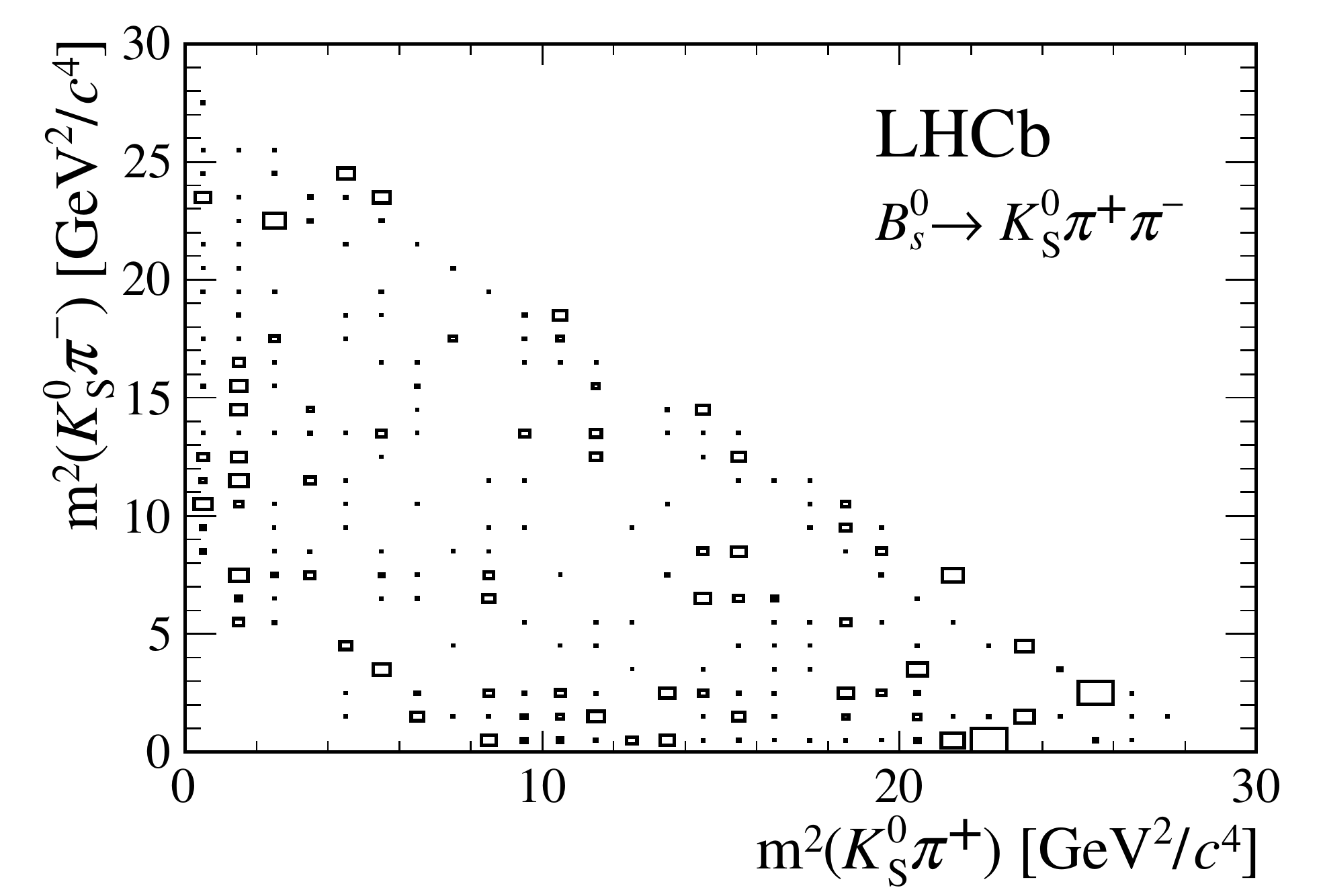}
\includegraphics[width=0.6\textwidth]{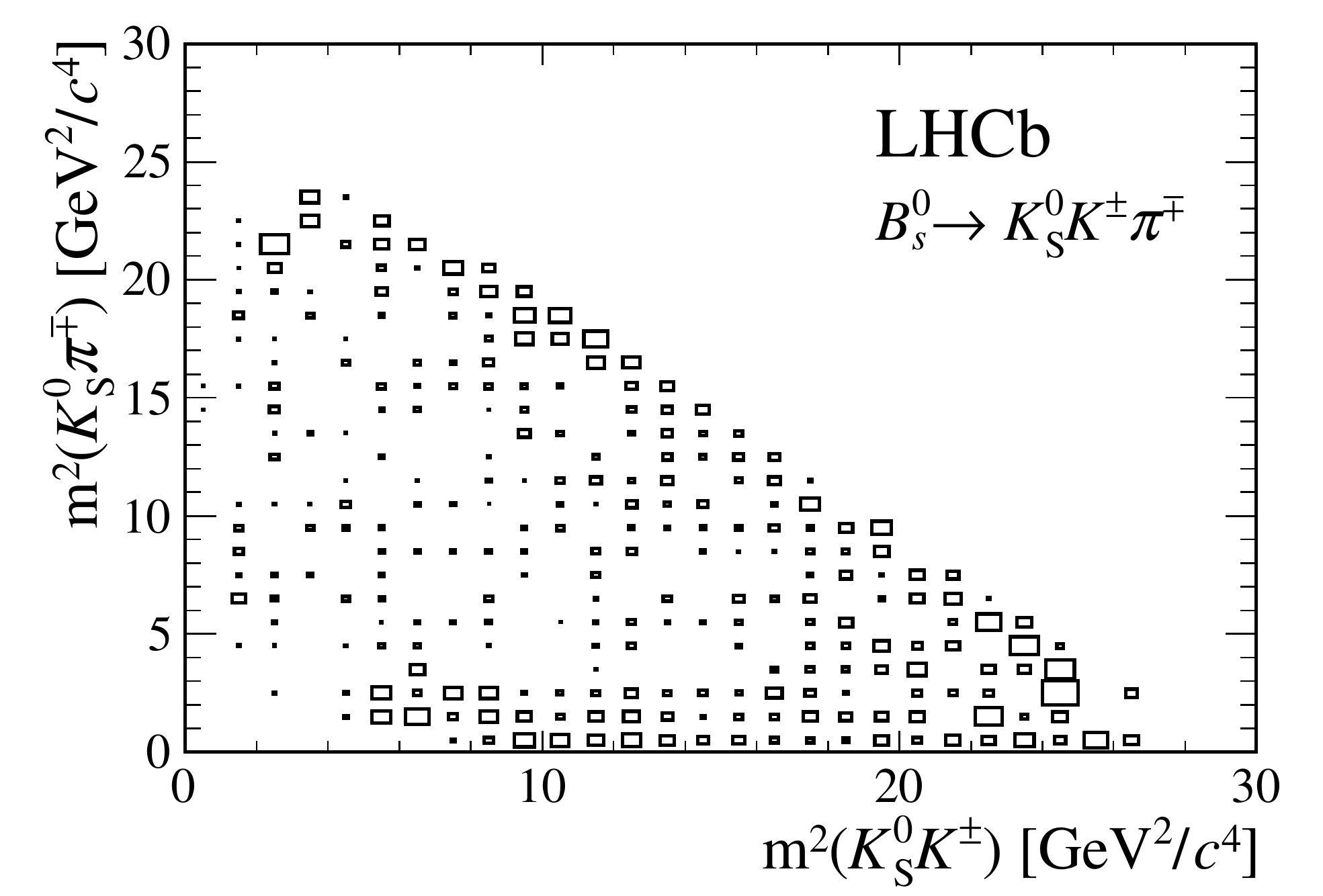}
\includegraphics[width=0.6\textwidth]{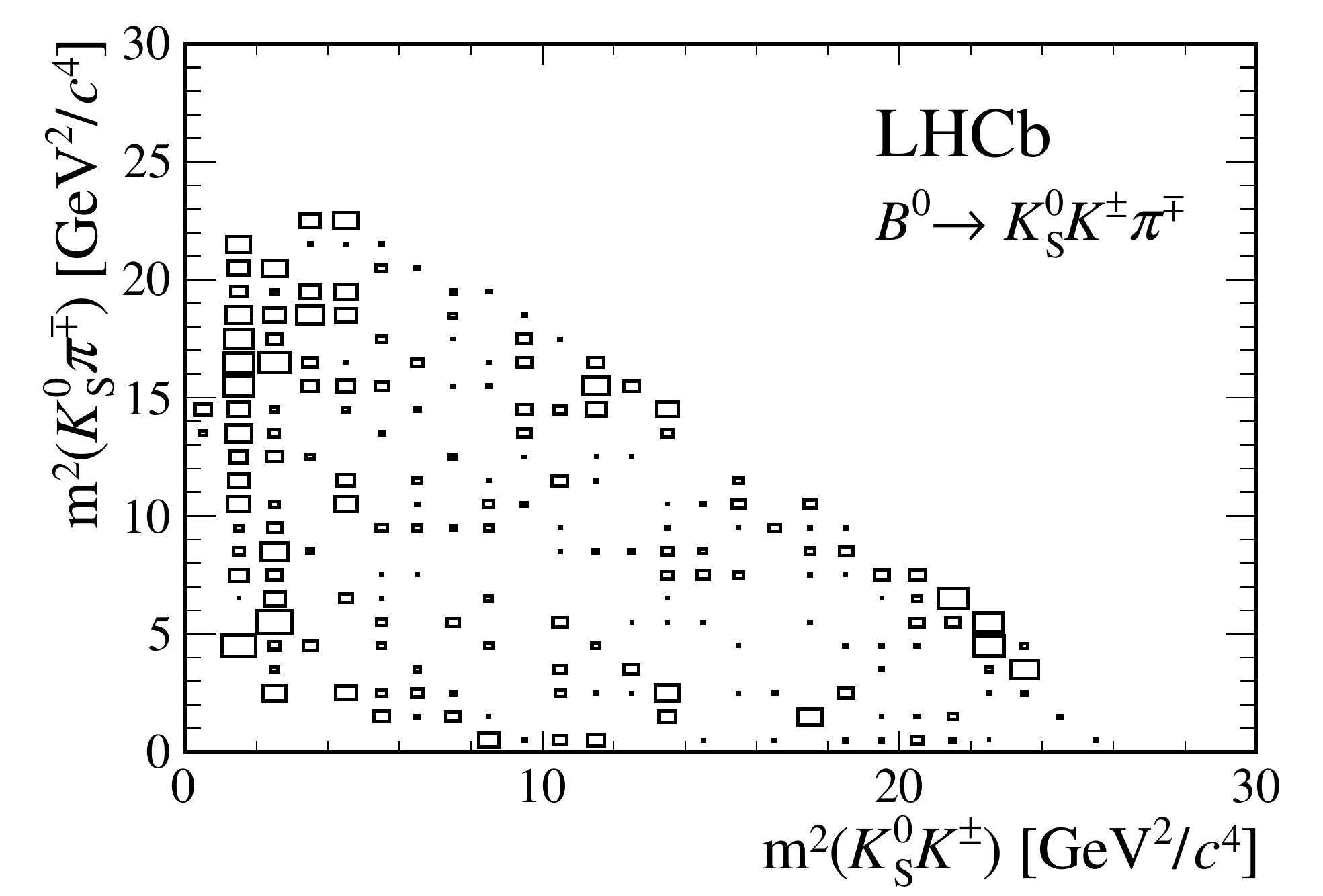}
}{
\includegraphics[width=0.6\textwidth]{figs/nBsKSpipi_All_splots_sw_bdtOBS_Dalitz-paper.eps}
\includegraphics[width=0.6\textwidth]{figs/nBsKSKpi_All_splots_sw_bdtBR_Dalitz-paper.eps}
\includegraphics[width=0.6\textwidth]{figs/nBdKSKpi_All_splots_sw_bdtOBS_Dalitz-paper.eps}
}
\caption{Efficiency-corrected Dalitz-plot distributions, produced using the
\sPlot\ procedure, of (top) \BstoKsPiPi, (middle) \BstoKsKPi and (bottom)
\BdtoKsKPi events.  Bins with negative content appear empty.}  
\label{fig : splot dalitz}
\end{center}
\end{figure}

Each branching fraction is measured (or limited) relative to that of \BdtoKsPiPi. 
The ratios of branching fractions are determined independently for the two \KS 
reconstruction categories and then combined by performing a weighted
average, excluding the uncertainty due to the ratio of hadronisation
fractions, since it is fully correlated between the two categories.
The \DD and \LL results all agree within two standard deviations, including
statistical and systematic uncertainties.
The results obtained from the combination are
\begin{eqnarray*}
\nonumber
\frac{\Br{\BdtoKsKPi}} {\Br{\BdtoKsPiPi}}  & = & 0.128 \pm 0.017 \; {\rm(stat.)} \; \pm 0.009 \; ({\rm syst.}) \,,\\ 
\nonumber                                              
\frac{\Br{\BdtoKsKK}}  {\Br{\BdtoKsPiPi}}  & = & 0.385 \pm 0.031 \; {\rm(stat.)} \; \pm 0.023 \; ({\rm syst.}) \,,\\
\nonumber                                              
\frac{\Br{\BstoKsPiPi}}{\Br{\BdtoKsPiPi}}  & = & 0.29\phantom{0} \pm 0.06\phantom{0} \; {\rm(stat.)} \; \pm  0.03\phantom{0} \; ({\rm syst.}) \; \pm 0.02\phantom{0} \; (f_s/f_d) \,, \\
\nonumber  
\frac{\Br{\BstoKsKPi}} {\Br{\BdtoKsPiPi}}  & = & 1.48\phantom{0} \pm 0.12\phantom{0} \; {\rm(stat.)} \; \pm  0.08\phantom{0} \; ({\rm syst.}) \; \pm 0.12\phantom{0} \; (f_s/f_d) \,,\\
\nonumber    
\frac{\Br{\BstoKsKK}}{\Br{\BdtoKsPiPi}}  &\in& [0.004;0.068]  \; {\rm at \;\; 90\% \; CL}  \,.
\end{eqnarray*}  

The measurement of the relative branching fractions of \BdtoKsKPi and
\BdtoKsKK are in good agreement with, and slightly more precise than, the
previous world average
results~\cite{Garmash:2003er,Garmash:2006fh,Aubert:2009me,delAmoSanchez:2010ur,Lees:2012kxa,HFAG,PDG2012}.
Using the world average value, ${\cal B}(\BdtoKzPiPi) = (4.96 \pm 0.20) \times 10^{-5}$~\cite{HFAG,PDG2012}, the measured time-integrated branching fractions
\begin{eqnarray}
\nonumber
\Br{\BdtoKzKPi}  &=& \phantom{0}(6.4 \pm 0.9  \pm  0.4 \pm 0.3)\times10^{-6} \,, \\
\nonumber
\Br{\BdtoKzKK}   &=& (19.1 \pm 1.5 \pm 1.1 \pm 0.8)\times10^{-6} \,,\\
\nonumber
\Br{\BstoKzPiPi} &=& (14.3 \pm 2.8 \pm 1.8 \pm 0.6)\times10^{-6} \,,\\
\nonumber
\Br{\BstoKzKPi}  &=& (73.6 \pm 5.7 \pm 6.9 \pm 3.0)\times10^{-6} \,,\\
\nonumber
\Br{\BstoKzKK}   &\in& [0.2;3.4] \times10^{-6} \; {\rm at \;\; 90\% \; CL} \,,
\end{eqnarray}  
are obtained, where the first uncertainty is statistical, the second 
systematic and the last due to the uncertainty on ${\cal B}(\BdtoKzPiPi)$.

The first observation of the decay modes \BstoKsPiPi and \BstoKsKPi is an
important step towards extracting information on the mixing-induced
\CP-violating phase in the \Bs system and the weak phase \Pgamma from these
decays.
The apparent rich structure of the Dalitz plots, particularly for the
\BdstoKsKPi decays, motivates future amplitude analyses of these
\BdstoKshhp modes with larger data samples.

\section*{Acknowledgements}

\noindent We express our gratitude to our colleagues in the CERN
accelerator departments for the excellent performance of the LHC. We
thank the technical and administrative staff at the LHCb
institutes. We acknowledge support from CERN and from the national
agencies: CAPES, CNPq, FAPERJ and FINEP (Brazil); NSFC (China);
CNRS/IN2P3 and Region Auvergne (France); BMBF, DFG, HGF and MPG
(Germany); SFI (Ireland); INFN (Italy); FOM and NWO (The Netherlands);
SCSR (Poland); MEN/IFA (Romania); MinES, Rosatom, RFBR and NRC
``Kurchatov Institute'' (Russia); MinECo, XuntaGal and GENCAT (Spain);
SNSF and SER (Switzerland); NAS Ukraine (Ukraine); STFC (United
Kingdom); NSF (USA). We also acknowledge the support received from the
ERC under FP7. The Tier1 computing centres are supported by IN2P3
(France), KIT and BMBF (Germany), INFN (Italy), NWO and SURF (The
Netherlands), PIC (Spain), GridPP (United Kingdom). We are thankful
for the computing resources put at our disposal by Yandex LLC
(Russia), as well as to the communities behind the multiple open
source software packages that we depend on.

\addcontentsline{toc}{section}{References}
\setboolean{inbibliography}{true}
\bibliographystyle{LHCb}
\bibliography{main,LHCb-PAPER,LHCb-CONF,LHCb-DP}

\ifx\mcitethebibliography\mciteundefinedmacro
\PackageError{LHCb.bst}{mciteplus.sty has not been loaded}
{This bibstyle requires the use of the mciteplus package.}\fi
\providecommand{\href}[2]{#2}
\begin{mcitethebibliography}{10}
\mciteSetBstSublistMode{n}
\mciteSetBstMaxWidthForm{subitem}{\alph{mcitesubitemcount})}
\mciteSetBstSublistLabelBeginEnd{\mcitemaxwidthsubitemform\space}
{\relax}{\relax}

\bibitem{Cabibbo:1963yz}
N.~Cabibbo, \ifthenelse{\boolean{articletitles}}{{\it Unitary symmetry and
  leptonic decays},
  }{}\href{http://dx.doi.org/10.1103/PhysRevLett.10.531}{Phys.\ Rev.\ Lett.\
  {\bf 10} (1963) 531}\relax
\mciteBstWouldAddEndPuncttrue
\mciteSetBstMidEndSepPunct{\mcitedefaultmidpunct}
{\mcitedefaultendpunct}{\mcitedefaultseppunct}\relax
\EndOfBibitem
\bibitem{Kobayashi:1973fv}
M.~Kobayashi and T.~Maskawa, \ifthenelse{\boolean{articletitles}}{{\it {\CP}
  violation in the renormalizable theory of weak interaction},
  }{}\href{http://dx.doi.org/10.1143/PTP.49.652}{Prog.\ Theor.\ Phys.\  {\bf
  49} (1973) 652}\relax
\mciteBstWouldAddEndPuncttrue
\mciteSetBstMidEndSepPunct{\mcitedefaultmidpunct}
{\mcitedefaultendpunct}{\mcitedefaultseppunct}\relax
\EndOfBibitem
\bibitem{Buchalla:2005us}
G.~Buchalla, G.~Hiller, Y.~Nir, and G.~Raz,
  \ifthenelse{\boolean{articletitles}}{{\it {The pattern of \CP asymmetries in
  \btos transitions}},
  }{}\href{http://dx.doi.org/10.1088/1126-6708/2005/09/074}{JHEP {\bf 09}
  (2005) 074}, \href{http://arxiv.org/abs/hep-ph/0503151}{{\tt
  arXiv:hep-ph/0503151}}\relax
\mciteBstWouldAddEndPuncttrue
\mciteSetBstMidEndSepPunct{\mcitedefaultmidpunct}
{\mcitedefaultendpunct}{\mcitedefaultseppunct}\relax
\EndOfBibitem
\bibitem{Grossman:1996ke}
Y.~Grossman and M.~P. Worah, \ifthenelse{\boolean{articletitles}}{{\it {\CP
  asymmetries in \B decays with new physics in decay amplitudes}},
  }{}\href{http://dx.doi.org/10.1016/S0370-2693(97)00068-3}{Phys.\ Lett.\  {\bf
  B395} (1997) 241}, \href{http://arxiv.org/abs/hep-ph/9612269}{{\tt
  arXiv:hep-ph/9612269}}\relax
\mciteBstWouldAddEndPuncttrue
\mciteSetBstMidEndSepPunct{\mcitedefaultmidpunct}
{\mcitedefaultendpunct}{\mcitedefaultseppunct}\relax
\EndOfBibitem
\bibitem{London:1997zk}
D.~London and A.~Soni, \ifthenelse{\boolean{articletitles}}{{\it {Measuring the
  \CP angle $\beta$ in hadronic \btos penguin decays}},
  }{}\href{http://dx.doi.org/10.1016/S0370-2693(97)00695-3}{Phys.\ Lett.\  {\bf
  B407} (1997) 61}, \href{http://arxiv.org/abs/hep-ph/9704277}{{\tt
  arXiv:hep-ph/9704277}}\relax
\mciteBstWouldAddEndPuncttrue
\mciteSetBstMidEndSepPunct{\mcitedefaultmidpunct}
{\mcitedefaultendpunct}{\mcitedefaultseppunct}\relax
\EndOfBibitem
\bibitem{Ciuchini:1997zp}
M.~Ciuchini {\em et~al.}, \ifthenelse{\boolean{articletitles}}{{\it {\CP
  violating \B decays in the Standard Model and Supersymmetry}},
  }{}\href{http://dx.doi.org/10.1103/PhysRevLett.79.978}{Phys.\ Rev.\ Lett.\
  {\bf 79} (1997) 978}, \href{http://arxiv.org/abs/hep-ph/9704274}{{\tt
  arXiv:hep-ph/9704274}}\relax
\mciteBstWouldAddEndPuncttrue
\mciteSetBstMidEndSepPunct{\mcitedefaultmidpunct}
{\mcitedefaultendpunct}{\mcitedefaultseppunct}\relax
\EndOfBibitem
\bibitem{Dalseno:2008wwa}
\belle collaboration, J.~Dalseno {\em et~al.},
  \ifthenelse{\boolean{articletitles}}{{\it {Time-dependent Dalitz-plot
  measurement of \CP parameters in \BdtoKsPiPi decays}},
  }{}\href{http://dx.doi.org/10.1103/PhysRevD.79.072004}{Phys.\ Rev.\  {\bf
  D79} (2009) 072004}, \href{http://arxiv.org/abs/0811.3665}{{\tt
  arXiv:0811.3665}}\relax
\mciteBstWouldAddEndPuncttrue
\mciteSetBstMidEndSepPunct{\mcitedefaultmidpunct}
{\mcitedefaultendpunct}{\mcitedefaultseppunct}\relax
\EndOfBibitem
\bibitem{Aubert:2009me}
\babar collaboration, B.~Aubert {\em et~al.},
  \ifthenelse{\boolean{articletitles}}{{\it {Time-dependent amplitude analysis
  of \BdtoKsPiPi}},
  }{}\href{http://dx.doi.org/10.1103/PhysRevD.80.112001}{Phys.\ Rev.\  {\bf
  D80} (2009) 112001}, \href{http://arxiv.org/abs/0905.3615}{{\tt
  arXiv:0905.3615}}\relax
\mciteBstWouldAddEndPuncttrue
\mciteSetBstMidEndSepPunct{\mcitedefaultmidpunct}
{\mcitedefaultendpunct}{\mcitedefaultseppunct}\relax
\EndOfBibitem
\bibitem{Nakahama:2010nj}
\belle collaboration, Y.~Nakahama {\em et~al.},
  \ifthenelse{\boolean{articletitles}}{{\it {Measurement of \CP violating
  asymmetries in \BdtoKsKK decays with a time-dependent Dalitz approach}},
  }{}\href{http://dx.doi.org/10.1103/PhysRevD.82.073011}{Phys.\ Rev.\  {\bf
  D82} (2010) 073011}, \href{http://arxiv.org/abs/1007.3848}{{\tt
  arXiv:1007.3848}}\relax
\mciteBstWouldAddEndPuncttrue
\mciteSetBstMidEndSepPunct{\mcitedefaultmidpunct}
{\mcitedefaultendpunct}{\mcitedefaultseppunct}\relax
\EndOfBibitem
\bibitem{Lees:2012kxa}
\babar Collaboration, J.~P. Lees {\em et~al.},
  \ifthenelse{\boolean{articletitles}}{{\it {Study of \CP violation in
  Dalitz-plot analyses of \BdtoKsKK, \decay{\Bp}{\Kp \Km \Kp}, and
  \decay{\Bp}{\KS \KS \Kp}}},
  }{}\href{http://dx.doi.org/10.1103/PhysRevD.85.112010}{Phys.\ Rev.\  {\bf
  D85} (2012) 112010}, \href{http://arxiv.org/abs/1201.5897}{{\tt
  arXiv:1201.5897}}\relax
\mciteBstWouldAddEndPuncttrue
\mciteSetBstMidEndSepPunct{\mcitedefaultmidpunct}
{\mcitedefaultendpunct}{\mcitedefaultseppunct}\relax
\EndOfBibitem
\bibitem{HFAG}
Heavy Flavor Averaging Group, Y.~Amhis {\em et~al.},
  \ifthenelse{\boolean{articletitles}}{{\it {Averages of $b$-hadron,
  $c$-hadron, and $\tau$-lepton properties as of early 2012}},
  }{}\href{http://arxiv.org/abs/1207.1158}{{\tt arXiv:1207.1158}}, {updated
  results and plots available at:
  \href{http://www.slac.stanford.edu/xorg/hfag/}{{\tt
  http://www.slac.stanford.edu/xorg/hfag/}}}\relax
\mciteBstWouldAddEndPuncttrue
\mciteSetBstMidEndSepPunct{\mcitedefaultmidpunct}
{\mcitedefaultendpunct}{\mcitedefaultseppunct}\relax
\EndOfBibitem
\bibitem{Silvestrini:2007yf}
L.~Silvestrini, \ifthenelse{\boolean{articletitles}}{{\it {Searching for new
  physics in $b \to s$ hadronic penguin decays}},
  }{}\href{http://dx.doi.org/10.1146/annurev.nucl.57.090506.123007}{Ann.\ Rev.\
  Nucl.\ Part.\ Sci.\  {\bf 57} (2007) 405},
  \href{http://arxiv.org/abs/0705.1624}{{\tt arXiv:0705.1624}}\relax
\mciteBstWouldAddEndPuncttrue
\mciteSetBstMidEndSepPunct{\mcitedefaultmidpunct}
{\mcitedefaultendpunct}{\mcitedefaultseppunct}\relax
\EndOfBibitem
\bibitem{Ciuchini:2006kv}
M.~Ciuchini, M.~Pierini, and L.~Silvestrini,
  \ifthenelse{\boolean{articletitles}}{{\it {New bounds on the CKM matrix from
  \decay{\B}{\kaon \pi \pi} Dalitz-plot analyses}},
  }{}\href{http://dx.doi.org/10.1103/PhysRevD.74.051301}{Phys.\ Rev.\  {\bf
  D74} (2006) 051301}, \href{http://arxiv.org/abs/hep-ph/0601233}{{\tt
  arXiv:hep-ph/0601233}}\relax
\mciteBstWouldAddEndPuncttrue
\mciteSetBstMidEndSepPunct{\mcitedefaultmidpunct}
{\mcitedefaultendpunct}{\mcitedefaultseppunct}\relax
\EndOfBibitem
\bibitem{Gronau:2006qn}
M.~Gronau, D.~Pirjol, A.~Soni, and J.~Zupan,
  \ifthenelse{\boolean{articletitles}}{{\it {Improved method for CKM
  constraints in charmless three-body B and \Bs decays}},
  }{}\href{http://dx.doi.org/10.1103/PhysRevD.75.014002}{Phys.\ Rev.\  {\bf
  D75} (2007) 014002}, \href{http://arxiv.org/abs/hep-ph/0608243}{{\tt
  arXiv:hep-ph/0608243}}\relax
\mciteBstWouldAddEndPuncttrue
\mciteSetBstMidEndSepPunct{\mcitedefaultmidpunct}
{\mcitedefaultendpunct}{\mcitedefaultseppunct}\relax
\EndOfBibitem
\bibitem{BABAR:2011ae}
\babar collaboration, J.~P. Lees {\em et~al.},
  \ifthenelse{\boolean{articletitles}}{{\it {Amplitude analysis of
  \decay{\Bd}{\Kp \pim \piz} and evidence of direct \CP violation in
  \decay{\B}{\Kstar \pi} decays}},
  }{}\href{http://dx.doi.org/10.1103/PhysRevD.83.112010}{Phys.\ Rev.\  {\bf
  D83} (2011) 112010}, \href{http://arxiv.org/abs/1105.0125}{{\tt
  arXiv:1105.0125}}\relax
\mciteBstWouldAddEndPuncttrue
\mciteSetBstMidEndSepPunct{\mcitedefaultmidpunct}
{\mcitedefaultendpunct}{\mcitedefaultseppunct}\relax
\EndOfBibitem
\bibitem{Ciuchini:2006st}
M.~Ciuchini, M.~Pierini, and L.~Silvestrini,
  \ifthenelse{\boolean{articletitles}}{{\it {Hunting the CKM weak phase with
  time-integrated Dalitz analyses of $\Bs \to K \pi \pi$ decays}},
  }{}\href{http://dx.doi.org/10.1016/j.physletb.2006.12.043}{Phys.\ Lett.\
  {\bf B645} (2007) 201}, \href{http://arxiv.org/abs/hep-ph/0602207}{{\tt
  arXiv:hep-ph/0602207}}\relax
\mciteBstWouldAddEndPuncttrue
\mciteSetBstMidEndSepPunct{\mcitedefaultmidpunct}
{\mcitedefaultendpunct}{\mcitedefaultseppunct}\relax
\EndOfBibitem
\bibitem{delAmoSanchez:2010ur}
\babar collaboration, P.~del Amo~Sanchez {\em et~al.},
  \ifthenelse{\boolean{articletitles}}{{\it {Observation of the rare decay
  \BdtoKsKPi}}, }{}\href{http://dx.doi.org/10.1103/PhysRevD.82.031101}{Phys.\
  Rev.\  {\bf D82} (2010) 031101}, \href{http://arxiv.org/abs/1003.0640}{{\tt
  arXiv:1003.0640}}\relax
\mciteBstWouldAddEndPuncttrue
\mciteSetBstMidEndSepPunct{\mcitedefaultmidpunct}
{\mcitedefaultendpunct}{\mcitedefaultseppunct}\relax
\EndOfBibitem
\bibitem{DeBruyn:2012wj}
K.~De~Bruyn {\em et~al.}, \ifthenelse{\boolean{articletitles}}{{\it {Branching
  ratio measurements of \Bs decays}},
  }{}\href{http://dx.doi.org/10.1103/PhysRevD.86.014027}{Phys.\ Rev.\  {\bf
  D86} (2012) 014027}, \href{http://arxiv.org/abs/1204.1735}{{\tt
  arXiv:1204.1735}}\relax
\mciteBstWouldAddEndPuncttrue
\mciteSetBstMidEndSepPunct{\mcitedefaultmidpunct}
{\mcitedefaultendpunct}{\mcitedefaultseppunct}\relax
\EndOfBibitem
\bibitem{Sjostrand:2006za}
T.~Sj\"{o}strand, S.~Mrenna, and P.~Skands,
  \ifthenelse{\boolean{articletitles}}{{\it {PYTHIA 6.4 physics and manual}},
  }{}\href{http://dx.doi.org/10.1088/1126-6708/2006/05/026}{JHEP {\bf 05}
  (2006) 026}, \href{http://arxiv.org/abs/hep-ph/0603175}{{\tt
  arXiv:hep-ph/0603175}}\relax
\mciteBstWouldAddEndPuncttrue
\mciteSetBstMidEndSepPunct{\mcitedefaultmidpunct}
{\mcitedefaultendpunct}{\mcitedefaultseppunct}\relax
\EndOfBibitem
\bibitem{LHCb-PROC-2010-056}
I.~Belyaev {\em et~al.}, \ifthenelse{\boolean{articletitles}}{{\it {Handling of
  the generation of primary events in \gauss, the \lhcb simulation framework}},
  }{}\href{http://dx.doi.org/10.1109/NSSMIC.2010.5873949}{Nuclear Science
  Symposium Conference Record (NSS/MIC) {\bf IEEE} (2010) 1155}\relax
\mciteBstWouldAddEndPuncttrue
\mciteSetBstMidEndSepPunct{\mcitedefaultmidpunct}
{\mcitedefaultendpunct}{\mcitedefaultseppunct}\relax
\EndOfBibitem
\bibitem{Lange:2001uf}
D.~J. Lange, \ifthenelse{\boolean{articletitles}}{{\it {The EvtGen particle
  decay simulation package}},
  }{}\href{http://dx.doi.org/10.1016/S0168-9002(01)00089-4}{Nucl.\ Instrum.\
  Meth.\  {\bf A462} (2001) 152}\relax
\mciteBstWouldAddEndPuncttrue
\mciteSetBstMidEndSepPunct{\mcitedefaultmidpunct}
{\mcitedefaultendpunct}{\mcitedefaultseppunct}\relax
\EndOfBibitem
\bibitem{Golonka:2005pn}
P.~Golonka and Z.~Was, \ifthenelse{\boolean{articletitles}}{{\it {PHOTOS Monte
  Carlo: a precision tool for QED corrections in $Z$ and $W$ decays}},
  }{}\href{http://dx.doi.org/10.1140/epjc/s2005-02396-4}{Eur.\ Phys.\ J.\  {\bf
  C45} (2006) 97}, \href{http://arxiv.org/abs/hep-ph/0506026}{{\tt
  arXiv:hep-ph/0506026}}\relax
\mciteBstWouldAddEndPuncttrue
\mciteSetBstMidEndSepPunct{\mcitedefaultmidpunct}
{\mcitedefaultendpunct}{\mcitedefaultseppunct}\relax
\EndOfBibitem
\bibitem{Allison:2006ve}
Geant4 collaboration, J.~Allison {\em et~al.},
  \ifthenelse{\boolean{articletitles}}{{\it {Geant4 developments and
  applications}}, }{}\href{http://dx.doi.org/10.1109/TNS.2006.869826}{IEEE
  Trans.\ Nucl.\ Sci.\  {\bf 53} (2006) 270}\relax
\mciteBstWouldAddEndPuncttrue
\mciteSetBstMidEndSepPunct{\mcitedefaultmidpunct}
{\mcitedefaultendpunct}{\mcitedefaultseppunct}\relax
\EndOfBibitem
\bibitem{Agostinelli:2002hh}
Geant4 collaboration, S.~Agostinelli {\em et~al.},
  \ifthenelse{\boolean{articletitles}}{{\it {Geant4: a simulation toolkit}},
  }{}\href{http://dx.doi.org/10.1016/S0168-9002(03)01368-8}{Nucl.\ Instrum.\
  Meth.\  {\bf A506} (2003) 250}\relax
\mciteBstWouldAddEndPuncttrue
\mciteSetBstMidEndSepPunct{\mcitedefaultmidpunct}
{\mcitedefaultendpunct}{\mcitedefaultseppunct}\relax
\EndOfBibitem
\bibitem{LHCb-PROC-2011-006}
M.~Clemencic {\em et~al.}, \ifthenelse{\boolean{articletitles}}{{\it {The \lhcb
  simulation application, \gauss: design, evolution and experience}},
  }{}\href{http://dx.doi.org/10.1088/1742-6596/331/3/032023}{{J.\ Phys.\ \!\!:
  Conf.\ Ser.\ } {\bf 331} (2011) 032023}\relax
\mciteBstWouldAddEndPuncttrue
\mciteSetBstMidEndSepPunct{\mcitedefaultmidpunct}
{\mcitedefaultendpunct}{\mcitedefaultseppunct}\relax
\EndOfBibitem
\bibitem{Alves:2008zz}
LHCb collaboration, A.~A. Alves~Jr. {\em et~al.},
  \ifthenelse{\boolean{articletitles}}{{\it {The \lhcb detector at the LHC}},
  }{}\href{http://dx.doi.org/10.1088/1748-0221/3/08/S08005}{JINST {\bf 3}
  (2008) S08005}\relax
\mciteBstWouldAddEndPuncttrue
\mciteSetBstMidEndSepPunct{\mcitedefaultmidpunct}
{\mcitedefaultendpunct}{\mcitedefaultseppunct}\relax
\EndOfBibitem
\bibitem{LHCb-DP-2012-003}
M.~Adinolfi {\em et~al.}, \ifthenelse{\boolean{articletitles}}{{\it
  {Performance of the \lhcb RICH detector at the LHC}},
  }{}\href{http://dx.doi.org/10.1140/epjc/s10052-013-2431-9}{Eur.\ Phys.\ J.\
  {\bf C73} (2013) 2431}, \href{http://arxiv.org/abs/1211.6759}{{\tt
  arXiv:1211.6759}}\relax
\mciteBstWouldAddEndPuncttrue
\mciteSetBstMidEndSepPunct{\mcitedefaultmidpunct}
{\mcitedefaultendpunct}{\mcitedefaultseppunct}\relax
\EndOfBibitem
\bibitem{LHCb-DP-2012-004}
R.~Aaij {\em et~al.}, \ifthenelse{\boolean{articletitles}}{{\it {The \lhcb
  trigger and its performance in 2011}},
  }{}\href{http://dx.doi.org/10.1088/1748-0221/8/04/P04022}{JINST {\bf 8}
  (2013) P04022}, \href{http://arxiv.org/abs/1211.3055}{{\tt
  arXiv:1211.3055}}\relax
\mciteBstWouldAddEndPuncttrue
\mciteSetBstMidEndSepPunct{\mcitedefaultmidpunct}
{\mcitedefaultendpunct}{\mcitedefaultseppunct}\relax
\EndOfBibitem
\bibitem{BBDT}
V.~V. Gligorov and M.~Williams, \ifthenelse{\boolean{articletitles}}{{\it
  {Efficient, reliable and fast high-level triggering using a bonsai boosted
  decision tree}},
  }{}\href{http://dx.doi.org/10.1088/1748-0221/8/02/P02013}{JINST {\bf 8}
  (2013) P02013}, \href{http://arxiv.org/abs/1210.6861}{{\tt
  arXiv:1210.6861}}\relax
\mciteBstWouldAddEndPuncttrue
\mciteSetBstMidEndSepPunct{\mcitedefaultmidpunct}
{\mcitedefaultendpunct}{\mcitedefaultseppunct}\relax
\EndOfBibitem
\bibitem{PDG2012}
Particle Data Group, J.~Beringer {\em et~al.},
  \ifthenelse{\boolean{articletitles}}{{\it {\href{http://pdg.lbl.gov/}{Review
  of particle physics}}},
  }{}\href{http://dx.doi.org/10.1103/PhysRevD.86.010001}{Phys.\ Rev.\  {\bf
  D86} (2012) 010001}\relax
\mciteBstWouldAddEndPuncttrue
\mciteSetBstMidEndSepPunct{\mcitedefaultmidpunct}
{\mcitedefaultendpunct}{\mcitedefaultseppunct}\relax
\EndOfBibitem
\bibitem{Breiman}
L.~Breiman, J.~H. Friedman, R.~A. Olshen, and C.~J. Stone, {\em Classification
  and regression trees}, Wadsworth international group, Belmont, California,
  USA, 1984\relax
\mciteBstWouldAddEndPuncttrue
\mciteSetBstMidEndSepPunct{\mcitedefaultmidpunct}
{\mcitedefaultendpunct}{\mcitedefaultseppunct}\relax
\EndOfBibitem
\bibitem{AdaBoost}
R.~E. Schapire and Y.~Freund, \ifthenelse{\boolean{articletitles}}{{\it A
  decision-theoretic generalization of on-line learning and an application to
  boosting}, }{}\href{http://dx.doi.org/10.1006/jcss.1997.1504}{Jour.\ Comp.\
  and Syst.\ Sc.\  {\bf 55} (1997) 119}\relax
\mciteBstWouldAddEndPuncttrue
\mciteSetBstMidEndSepPunct{\mcitedefaultmidpunct}
{\mcitedefaultendpunct}{\mcitedefaultseppunct}\relax
\EndOfBibitem
\bibitem{Punzi:2003bu}
G.~Punzi, \ifthenelse{\boolean{articletitles}}{{\it {Sensitivity of searches
  for new signals and its optimization}}, }{}eConf {\bf C030908} (2003)
  MODT002, \href{http://arxiv.org/abs/physics/0308063}{{\tt
  arXiv:physics/0308063}}\relax
\mciteBstWouldAddEndPuncttrue
\mciteSetBstMidEndSepPunct{\mcitedefaultmidpunct}
{\mcitedefaultendpunct}{\mcitedefaultseppunct}\relax
\EndOfBibitem
\bibitem{Skwarnicki:1986xj}
T.~Skwarnicki, {\em {A study of the radiative cascade transitions between the
  Upsilon-prime and Upsilon resonances}}, PhD thesis, Institute of Nuclear
  Physics, Krakow, 1986,
  {\href{http://inspirehep.net/record/230779/files/230779.pdf}{DESY-F31-86-02}}\relax
\mciteBstWouldAddEndPuncttrue
\mciteSetBstMidEndSepPunct{\mcitedefaultmidpunct}
{\mcitedefaultendpunct}{\mcitedefaultseppunct}\relax
\EndOfBibitem
\bibitem{Albrecht:1990cs}
ARGUS collaboration, H.~Albrecht {\em et~al.},
  \ifthenelse{\boolean{articletitles}}{{\it {Exclusive hadronic decays of \B
  mesons}}, }{}\href{http://dx.doi.org/10.1007/BF01614687}{Z.\ Phys.\  {\bf
  C48} (1990) 543}\relax
\mciteBstWouldAddEndPuncttrue
\mciteSetBstMidEndSepPunct{\mcitedefaultmidpunct}
{\mcitedefaultendpunct}{\mcitedefaultseppunct}\relax
\EndOfBibitem
\bibitem{LHCb-PAPER-2012-037}
LHCb collaboration, R.~Aaij {\em et~al.},
  \ifthenelse{\boolean{articletitles}}{{\it {Measurement of the fragmentation
  fraction ratio $f_s/f_d$ and its dependence on $B$ meson kinematics}},
  }{}\href{http://dx.doi.org/10.1007/JHEP04(2013)001}{JHEP {\bf 04} (2013) 1},
  \href{http://arxiv.org/abs/1301.5286}{{\tt arXiv:1301.5286}}\relax
\mciteBstWouldAddEndPuncttrue
\mciteSetBstMidEndSepPunct{\mcitedefaultmidpunct}
{\mcitedefaultendpunct}{\mcitedefaultseppunct}\relax
\EndOfBibitem
\bibitem{Dalitz:1953cp}
R.~H. Dalitz, \ifthenelse{\boolean{articletitles}}{{\it {On the analysis of
  tau-meson data and the nature of the tau-meson}},
  }{}\href{http://dx.doi.org/10.1080/14786441008520365}{Phil.\ Mag.\  {\bf 44}
  (1953) 1068}\relax
\mciteBstWouldAddEndPuncttrue
\mciteSetBstMidEndSepPunct{\mcitedefaultmidpunct}
{\mcitedefaultendpunct}{\mcitedefaultseppunct}\relax
\EndOfBibitem
\bibitem{Aubert:2005sk}
\babar collaboration, B.~Aubert {\em et~al.},
  \ifthenelse{\boolean{articletitles}}{{\it {An amplitude analysis of the decay
  $B^\pm \to \pi^\pm \pi^\pm \pi^\mp$}},
  }{}\href{http://dx.doi.org/10.1103/PhysRevD.72.052002}{Phys.\ Rev.\  {\bf
  D72} (2005) 052002}, \href{http://arxiv.org/abs/hep-ex/0507025}{{\tt
  arXiv:hep-ex/0507025}}\relax
\mciteBstWouldAddEndPuncttrue
\mciteSetBstMidEndSepPunct{\mcitedefaultmidpunct}
{\mcitedefaultendpunct}{\mcitedefaultseppunct}\relax
\EndOfBibitem
\bibitem{Pivk:2004ty}
M.~Pivk and F.~R. Le~Diberder, \ifthenelse{\boolean{articletitles}}{{\it
  {sPlot: a statistical tool to unfold data distributions}},
  }{}\href{http://dx.doi.org/10.1016/j.nima.2005.08.106}{Nucl.\ Instrum.\
  Meth.\  {\bf A555} (2005) 356},
  \href{http://arxiv.org/abs/physics/0402083}{{\tt
  arXiv:physics/0402083}}\relax
\mciteBstWouldAddEndPuncttrue
\mciteSetBstMidEndSepPunct{\mcitedefaultmidpunct}
{\mcitedefaultendpunct}{\mcitedefaultseppunct}\relax
\EndOfBibitem
\bibitem{Feldman:1997qc}
G.~J. Feldman and R.~D. Cousins, \ifthenelse{\boolean{articletitles}}{{\it {A
  unified approach to the classical statistical analysis of small signals}},
  }{}Phys.\ Rev.\  {\bf D57} (1998) 3873\relax
\mciteBstWouldAddEndPuncttrue
\mciteSetBstMidEndSepPunct{\mcitedefaultmidpunct}
{\mcitedefaultendpunct}{\mcitedefaultseppunct}\relax
\EndOfBibitem
\bibitem{Garmash:2003er}
\belle collaboration, A.~Garmash {\em et~al.},
  \ifthenelse{\boolean{articletitles}}{{\it {Study of \B meson decays to three
  body charmless hadronic final states}},
  }{}\href{http://dx.doi.org/10.1103/PhysRevD.69.012001}{Phys.\ Rev.\  {\bf
  D69} (2004) 012001}, \href{http://arxiv.org/abs/hep-ex/0307082}{{\tt
  arXiv:hep-ex/0307082}}\relax
\mciteBstWouldAddEndPuncttrue
\mciteSetBstMidEndSepPunct{\mcitedefaultmidpunct}
{\mcitedefaultendpunct}{\mcitedefaultseppunct}\relax
\EndOfBibitem
\bibitem{Garmash:2006fh}
Belle Collaboration, A.~Garmash {\em et~al.},
  \ifthenelse{\boolean{articletitles}}{{\it {Dalitz analysis of three-body
  charmless $\Bz \to \Kz \pip \pim$ decay}},
  }{}\href{http://dx.doi.org/10.1103/PhysRevD.75.012006}{Phys.\ Rev.\  {\bf
  D75} (2007) 012006}, \href{http://arxiv.org/abs/hep-ex/0610081}{{\tt
  arXiv:hep-ex/0610081}}\relax
\mciteBstWouldAddEndPuncttrue
\mciteSetBstMidEndSepPunct{\mcitedefaultmidpunct}
{\mcitedefaultendpunct}{\mcitedefaultseppunct}\relax
\EndOfBibitem
\end{mcitethebibliography}

\end{document}